\documentclass[10pt,a4paper]{article}
\usepackage{amsmath}
\usepackage{amssymb}
\usepackage{array}
\usepackage{calc}
\usepackage{longtable}
\usepackage{multirow,booktabs}
\usepackage{cite,mcite}
\usepackage{relsize}
\usepackage{graphicx}
\usepackage{xspace}
\usepackage{units}
\usepackage{afterpage}
\usepackage{placeins}

\numberwithin{equation}{section}
\usepackage{mciteplus}
\usepackage[pdfborder={0 0 0}]{hyperref}
\usepackage[format=hang,labelfont=bf,hypcap=true]{caption}
\usepackage{subcaption}
\usepackage{sectsty}
\allsectionsfont{\sffamily}
\subsubsectionfont{\mdseries\itshape\large}
\makeatletter
\DeclareRobustCommand*{\bfseries}{%
  \not@math@alphabet\bfseries\mathbf
  \fontseries\bfdefault\selectfont
  \boldmath
}
\makeatother
\let\spreprint\empty

\let\sinstitute\empty

\makeatletter
\renewcommand{\maketitle}{\begingroup
  \null\thispagestyle{empty}%
    \ifx\spreprint\empty
      \vskip 5ex
    \else
      \flushright\large\spreprint\vskip 2ex
    \fi
    \vskip 5ex
    \flushleft
      {\sffamily\bfseries\huge\@title}\vskip 6ex
      \@author\vskip 2ex
      \ifx\sinstitute\empty
      \else
        {\small\sinstitute}
      \fi
    \vskip 5ex
  \endgroup
}
\makeatother
\renewenvironment{abstract}{\begin{center}
  {\large\sffamily\bfseries Abstract: }
  \begin{minipage}[t]{0.75\textwidth}
}{\end{minipage}\end{center}\vskip 10ex}

\numberwithin{equation}{section}

\newcommand{\bea}{\begin{eqnarray}}
\newcommand{\eea}{\end{eqnarray}}
\newcommand{\beq}{\begin{equation}}
\newcommand{\eeq}{\end{equation}}

\newcommand{\bs}{\begin{split}}
\newcommand{\es}{\end{split}}

\newcommand{\bi}{\begin{itemize}}
\newcommand{\ei}{\end{itemize}}
\newcommand{\bc}{\begin{center}}
\newcommand{\ec}{\end{center}}
\newcommand{\bac}{\begin{array}{c}}
\newcommand{\bacc}{\begin{array}{cc}}
\newcommand{\ea}{\end{array}}

\def\spa#1.#2{\langle#1\,#2\rangle}
\def\spb#1.#2{[#1\,#2]}

\newcommand{\sla}[1]{\ensuremath{{#1\kern-0.45em/}}}

\newcommand{\ie}{{\itshape i.e.}\xspace}

\newcommand\lep{\texorpdfstring{L\scalebox{0.8}{EP}\xspace}{LEP\xspace}}

\newcommand\LEP{\lep}

\newcommand\HERA{\texorpdfstring{H\scalebox{0.8}{ERA}\xspace}{HERA\xspace}}
\newcommand\EIC{\texorpdfstring{E\scalebox{0.8}{IC}\xspace}{EIC\xspace}}

\newcommand\OPAL{\opal}
\newcommand\opal{O\scalebox{0.8}{PAL}\xspace}

\newcommand\hera{H\scalebox{0.8}{ERA}\xspace}

\newcommand\zeus{Z\scalebox{0.8}{EUS}\xspace}
\newcommand\ZEUS{\zeus}

\newcommand{\MCatNLO}{M\protect\scalebox{0.8}{C}@N\protect\scalebox{0.8}{LO}\xspace}

\newcommand{\Herwig}{\texorpdfstring{H\protect\scalebox{0.8}{ERWIG}\xspace}{HERWIG\xspace}}

\newcommand{\Pythia}{\texorpdfstring{P\protect\scalebox{0.8}{YTHIA}\xspace}{PYTHIA\xspace}}

\newcommand{\OpenLoops}{O\protect\scalebox{0.8}{PEN}L\protect\scalebox{0.8}{OOPS}\xspace}

\newcommand{\Sherpa}{\texorpdfstring{S\protect\scalebox{0.8}{HERPA}\xspace}{SHERPA\xspace}}
\newcommand{\Comix}{C\protect\scalebox{0.8}{OMIX}\xspace}

\newcommand{\Amegic}{A\protect\scalebox{0.8}{MEGIC}\xspace}
\newcommand{\CSS}{CSS\protect\scalebox{0.8}{HOWER}\xspace}

\newcommand{\Ahadic}{A\protect\scalebox{0.8}{HADIC}\xspace}

\newcommand{\Rivet}{R\protect\scalebox{0.8}{IVET}\xspace}

\usepackage{authblk}
\usepackage[symbol]{footmisc}
\hypersetup{
  pdfauthor={Ilkka Helenius, Peter Meinzinger, Simon Plätzer, Peter Richardson}
  pdftitle={Photoproduction in general-purpose event generators},
  pdfkeywords={QCD, Event Generators, Photon PDF, HERA, LEP}
}

\begin{document}
\title{Photoproduction in general-purpose event generators}
\author[1,2]{I.~Helenius\footnote[2]{ilkka.m.helenius@jyu.fi}}
\author[3*]{P.~Meinzinger\footnote[3]{peter.meinzinger@durham.ac.uk}}
\author[4,5]{S.~Plätzer\footnote[4]{simon.plaetzer@uni-graz.at}}
\author[3]{P.~Richardson\footnote[5]{peter.richardson@durham.ac.uk}}
\affil[1]{University of Jyvaskyla, Department of Physics, P.O. Box 35, FI-40014 University of Jyvaskyla,
Finland}
\affil[2]{Helsinki Institute of Physics, P.O. Box 64, FI-00014 University of Helsinki, Finland}
\affil[3]{Institute for Particle Physics Phenomenology, Durham University, Durham DH1 3LE, UK}
\affil[4]{Institut f\"ur Physik, NAWI, Universit\"at Graz, Universit\"atsplatz 5, 8010 Graz, AT}\affil[5]{Teilchenphysik, Fakult\"at f\"ur Physik, Universit\"at Wien, Boltzmanngasse 5, 1030 Wien, AT}
\date{\today}

\maketitle

\begin{abstract}
We compare the three general-purpose Monte Carlo event generators, \Herwig, \Pythia, and \Sherpa for jet photoproduction processes in $e^+e^-$ and $ep$ collisions.
Due to the lower energy scales probed, photoproduction is particularly sensitive to non-perturbative corrections. In a systematic analysis we disentangle and quantify the differences between the generators in these processes, \ie contributions from beam remnants, parton showers, multiparton interactions (MPIs), and hadronisation modelling.
We outline the default inputs and implementation differences and compare the computations with experimental data from \lep and \hera. We find that all generators provide a decent description of the data within the uncertainties, with particularly good descriptions by the LO-accurate \Pythia and the NLO-accurate \Sherpa. Finally, we also present predictions for the upcoming EIC for jet observables and event shapes and conclude that a modern global refit of the photon parton distributions and dedicated experimental measurements ported to the \Rivet{} framework to constrain non-perturbative parameters are the
key prerequisites for precision photoproduction phenomenology at the \EIC{}.
\end{abstract}

\section{Introduction}
General-purpose Monte Carlo (MC) event generators \cite{Seymour:2013ega} are computational tools for high-energy collisions, providing complete simulations of events for various beam configurations.
Their modelling covers a broad range of energy scales, from the hard partonic scattering down to hadronisation and decays of unstable hadrons.
The resulting distributions can be compared to data from experimental measurements and hence connect theoretical first-principle calculations to the experiments.
As a prime example, virtually all analyses published by the respective collaborations at the LHC use these MC simulations as part of their pipeline.

The main players in the field of general-purpose event generators are \Herwig \cite{Bahr:2008pv, Bellm:2015jjp}, \Pythia \cite{Bierlich:2022pfr} and \Sherpa \cite{Gleisberg:2008ta,Bothmann:2019yzt}, each with a decades-long development history.
The theoretical framework that they are built upon is collinear factorization and perturbative Quantum Chromodynamics (pQCD) \cite{Collins:1989gx}, allowing them to calculate differential cross-sections of varying perturbative order for relevant partonic scatterings.
Furthermore, the radiation of the initial and final partons participating in the hard scattering can be modelled using
parton-shower algorithms \cite{Hoche:2014rga} that turn the inclusive evolution equations into exclusive emission probabilities using Sudakov factors.
In the case of beam particles with partonic structure, the Underlying Event, i.e. additional subdominant scatterings, are modelled in multiparton interactions (MPIs) \cite{Bartalini:2018qje} and the conservation of quantum numbers and momentum is modelled as beam remnants.
Finally, the colour-charged partons are turned into colour-neutral particles through non-perturbative hadronization \cite{Andersson:1983ia, Webber:1983if} and colour reconnection models \cite{Sjostrand:1993hi, Christiansen:2015yqa}.

The development of the current event generators has been largely driven by the studies of proton-proton (pp) collisions at the LHC.
There, the high luminosities have enabled measurements to reach accuracies of a few percent-level, which necessitated theory predictions to reach matching or even higher accuracies.
For example, algorithms matching parton showers consistently with matrix-element computations at next-to-leading order (NLO) \cite{Frixione:2002ik, Frixione:2007vw} have been developed and implemented in all major event generators, for some particular cases even beyond NLO \cite{Lavesson:2008ah, Hoche:2018gti, Bertone:2022hig}.
Apart from the LHC, the modern event generators have been extended to handle other beam setups including heavy-ion collisions \cite{Bierlich:2018xfw, Helenius:2024vdj} and electron-proton (ep) collisions \cite{Cabouat:2017rzi, Hoche:2018gti, Helenius:2019gbd, Hoeche:2023gme, Knobbe:2023ehi}.
For the latter, the main driver is the Electron-Ion Collider (\EIC), the next large high-energy collider project at Brookhaven National Laboratory in US \cite{AbdulKhalek:2021gbh}.

Electron-proton collisions can be classified in terms of $Q^2$, the virtuality of the exchanged photon.
In deep inelastic scattering processes (DIS) the exchanged photon is far off-shell; however, in the photoproduction limit, $Q^2 \rightarrow 0$, the photon is almost on-shell and may fluctuate into a hadronic state \cite{Bauer:1977iq, Schuler:1992dt, Butterworth:2005aq}. In this case, the underlying hard scattering takes place between two coloured partons and is thus similar to a collision between two hadrons \cite{Klasen:2002xb}. In fact, this resolved component can dominate the total cross section of the photon-proton ($\gamma$p) system and has to be accounted for in addition to contributions of a direct photon scattering.
Following a similar line of argument, one can study photon-photon ($\gamma \gamma$) collisions at electron-positron (e$^+$e$^-$) colliders \cite{Walsh:1973mz, Frazer:1979gc, Schuler:1996en}.

Photoproduction has not seen much attention in recent decades, but predictions for these events are needed for the precision targets at the \EIC.
This study compares the three general purpose event generators, \Pythia, \Herwig and \Sherpa, with respect to photoproduction.
In Sec.~\ref{sec:MCs} we present the frameworks for each of the generators and discuss their differences.
To quantify their differences, we conduct a step-by-step analysis in Sec.~\ref{sec:diffs}.
In Sec.~\ref{sec:lep} and~\ref{sec:hera} we present cross sections for dijet photoproduction and compare the results with the \lep and \hera data.
We also show predictions for dijet photoproduction at kinematics relevant to the EIC including comparisons to event-shape and inclusive QCD observables in Sec.~\ref{sec:eic}.
We conclude our findings and provide a brief outlook in Sec.~\ref{sec:conclusion}.

\section{Event Generators}\label{sec:MCs}

For photons resolved into a hadronic state, the partonic structure can be described with DGLAP-evolved parton distribution functions (PDFs), $f^{\,\gamma}_i(x_{\gamma}, \mu_\mathrm{F}^2)$, where $x_{\gamma}$ is the momentum fraction of a given parton with respect to the photon and $\mu_\mathrm{F}^2$ is the factorization scale at which the PDFs are evaluated. Similarly as the PDFs for protons, $f^{\, \mathrm{p}}_i(x_{\mathrm{p}}, \mu_\mathrm{F}^2)$, the photon PDFs are obtained in a global analysis where the parameters of a non-perturbative input are fitted to experimental data \cite{Gluck:1983bh, Gluck:1991jc, Abramowicz:1993xb, Nisius:1999cv, Krawczyk:2000mf}. In addition to a fitted hadron-like part, the photon PDFs contain also a point-like (anomalous) contribution from perturbative $\gamma \rightarrow q \overline{q}$ splittings \cite{DeWitt:1978wn} that should be included in the evolution equations when comparing to experimental data. The structure of the resolved photons were an active topic around 20 years ago when LEP was producing experimental data for such analyses but, since then, activity has been sparse and the available fits are lacking the recent methodological developments, e.g. for the uncertainty estimation.

Knowing the structure of the incoming states, one can factorize hard-process generation for photoproduction in $ep$ collisions as
\begin{equation}
\mathrm{d} \sigma^{\mathrm{ep}}_{\mathrm{hard}} = f^{\,\mathrm{e}}_{\gamma}(x, Q^2) \otimes f_i^{\gamma}(x_{\gamma}, \mu_\mathrm{F}^2) \otimes f_j^{\mathrm{p}}(x_\mathrm{p}, \mu_\mathrm{F}^2) \otimes \mathrm{d} \sigma^{i j}_{\mathrm{hard}},
\label{eq:sigmahardEP}
\end{equation}
where $f^{\,\mathrm{e}}_{\gamma}(x, Q^2)$ is the photon flux from an electron at a given momentum fraction $x$ and photon virtuality $Q^2$, and $\sigma^{i j}_{\mathrm{hard}}$ perturbatively calculable hard coefficient function for a given hard process with initiating particles $i$ and $j$.
In case of direct-photon scattering the photon will act as an initiating particle and there one can replace the $f_i^{\gamma}$ with a delta function.
Thus, there are two contributions that should be accounted: direct and resolved.
At leading order (LO) these contributions can be taken as separate but at higher orders one has to make sure that no double counting takes place when combining these two event classes.
The photon flux can be computed using equivalent photon approximation (EPA) \cite{Budnev:1975poe}.
For photons from a charged lepton $l$ it gives in the virtuality-integrated leading-log (LL) approximation
\begin{equation}
f^{\,l}_{\gamma}(x) = \frac{\alpha_{\mathrm{em}}}{2 \pi} \frac{1 + (1-x)^2}{x} \log \left[ \frac{Q^2_{\mathrm{max}}} {Q^2_{\mathrm{min}}(x, m_l)} \right].
\label{eq:fluxLL}
\end{equation}
The lower limit for the virtuality follows from kinematics and the upper limit is adjusted according to the experimental configuration.
One can also account for the non-logarithmic (NL) term in the photon flux to improve the precision of the flux calculation. This correction can be written as~\cite{Frixione:1993yw}
\begin{equation}
    f_\mathrm{NL}(x) = - \frac{\alpha_\mathrm{em}}{\pi} m_l^2 x \left(\frac{1}{Q^2_\mathrm{min}(x, m_l)} - \frac{1}{Q^2_\mathrm{max}}\right) \ .
\label{eq:fluxNL}
\end{equation}
The correction comes with an overall negative sign and hence will lead to smaller total cross-sections. While it is seemingly a small correction in case of an electron beam due to its low mass, one should keep in mind that $Q^2_\mathrm{min}(x, m_l) \propto m_l^2$, so the mass dependence cancels out and the NL correction may indeed be relevant at kinematics where the photons are probed at large-$x$.

In case of $\gamma \gamma$ collisions in e$^+$e$^-$ the corresponding factorization reads
\begin{equation}
\mathrm{d} \sigma^{\mathrm{ee}}_{\mathrm{hard}} = f^{\,\mathrm{e}}_{\gamma}(x_1, Q_1^2) \otimes f_i^{\gamma}(x^{+}_{\gamma}, \mu_\mathrm{F}^2) \otimes f^{\,\mathrm{e}}_{\gamma}(x_2, Q_2^2) \otimes f_j^{\gamma}(x^{-}_{\gamma}, \mu_\mathrm{F}^2) \otimes \mathrm{d} \sigma^{i j}_{\mathrm{hard}}.
\label{eq:sigmahardEE}
\end{equation}
In this case there are four different contributions, direct-direct, direct-resolved, resolved-direct and resolved-resolved that need to be accounted for.
This factorized approach provides the starting point for all considered event generators over which further modelling of parton showers, beam remnants, MPIs and hadronization are included.
Thus, even though the starting point is the same for all generators, the results will depend on the applied inputs and particular choices made in this modelling.
Before comparing the different event generators to data, we will therefore describe the main features of the generators and the default inputs, and highlight some significant differences in the following subsections.

\subsection{\protect\Herwig}

\Herwig \cite{Corcella:2000bw,Bahr:2008pv,Bellm:2015jjp,Bewick:2023tfi} is a multi-purpose event generator, which has traditionally been focusing on accurate QCD simulations.
It includes two parton shower modules, an angular ordered one based on coherent branching \cite{Gieseke:2003rz}, and more recently a dipole parton shower \cite{Platzer:2009jq,Platzer:2011bc}.
Both parton shower modules include full mass effects, spin correlations, and initial and final state radiation \cite{Bellm:2017idv,Richardson:2018pvo}.
Hard processes can be simulated using a wide range of built-in matrix elements at leading and next to leading order, and via the Matchbox module \cite{Platzer:2011bc,Bellm:2015jjp} using external matrix element providers and an automated matching to NLO QCD either within the MC@NLO or POWHEG approaches.
Multi-jet merging \cite{Platzer:2012bs,Bellm:2017ktr} is available using the dipole shower.
Photoproduction processes can be simulated in e$^+$e$^-$, ep and pp collisions including all direct and resolved components.
The PDF sets and fluxes we currently provide are the SaS photon PDFs \cite{Schuler:1994ft} with the 1D, 1M, 2D and 2M schemes, and photon fluxes following a Weizs\"acker-Williams parametrization for electrons, and the Budnev approach \cite{Budnev:1974de} for protons.
Hadronization is modelled using the cluster hadronization model \cite{Bahr:2008pv}, and we provide an eikonal MPI model, which currently however is limited to $pp$ collisions and unfortunately cannot yet treat resolved photons due to technical issues.
Similar restrictions currently apply to NLO matching using Matchbox, though both of these issues are currently being worked on and should be available with one of the next \Herwig releases.
The \Herwig parton showers always need to terminate on a valence quark of the incoming parton, and thus create a remnant which complements the parton which had been extracted.
In case of the photon we will therefore always generate an anti-quark remnant, which might cause additional (collinear) shower activity even if radiation in the showers is switched off entirely.
Scale variations are not included in the \Herwig predictions presented in this study.

\subsection{\protect\Pythia}

Simulations for photoproduction in ep and $\gamma \gamma$ collisions in e$^+$e$^-$ have been fully enabled from release 8.226 onwards and the current setup is described in the 8.3 manual (Ref.~\cite{Bierlich:2022pfr}). Both hard and soft QCD processes can be generated, the former including processes like inclusive and diffractive jet production, and the latter non-diffractive and diffractive events and also elastic scattering. The implementation of resolved processes is based on CJKL photon PDFs~\cite{Cornet:2002iy} which includes hadron-like part with non-perturbative input separately from the perturbatively calculated anomalous contribution. In the generation of resolved processes these are considered simultaneously but the parton-shower algorithm includes the $\gamma \rightarrow q \overline{q}$ splitting that may collapse the resolved photon into an unresolved state at a perturbative scale corresponding to the anomalous part of the real-photon DGLAP evolution equation.

The simulations include generation of MPIs as long as the photons are in a resolved state and the MPI framework, where the QCD cross sections are regulated with the screening parameter $p_{\mathrm{T,0}}$, can be applied also for non-diffractive soft QCD processes without any phase-space cuts. The cross section for the different soft processes are obtained from the SaS parametrizations in Ref.~\cite{Schuler:1994ft}. For proton PDFs the current default NNPDF23\_lo\_as\_0130\_qed has been applied and the value of $\alpha_{\mathrm{S}}$ has been fixed accordingly to 0.130 at $\mu_{\mathrm{R}}^2 = M_{\mathrm{Z}}^2$. Here we have used \Pythia version 8.310 released in July 2023.

In this study we have applied the default setup for photoproduction in \Pythia 8.3. In case of $\gamma \gamma$ a specific tune for $p_{\mathrm{T,0}}$ has been applied based on single-inclusive charged-particle production data from OPAL at LEP \cite{OPAL:2006pyk} and in case $\gamma$p the $p_{\mathrm{T,0}}^{\mathrm{ref}}$ in the standard parametrization has been adjusted to 3.0~GeV based on comparisons to single-inclusive charged particle production in photoproduction at HERA~\cite{Helenius:2017aqz} which is in line also with the multiplicity distributions measured recently by ZEUS in Ref.~\cite{ZEUS:2021qzg}. Heavy quark masses are included apart from the $g Q \rightarrow g Q$ subprocess where massless expressions for the matrix element are applied also for charm and bottom quarks. The remnants are constructed by adding a minimal number of partons such that momentum and colour are conserved. In case of resolved photons the flavour for ``valence'' quark-antiquark pair is sampled according to relative fractions of the PDFs at the specific kinematics and a minimal fraction of the phase-space is cut out to allow room for the remnant partons. The scale variations have been performed by varying renormalization and factorization scales independently in the hard process generation by a factor of two, excluding the variations where the relative difference would be 4.

\subsection{\protect\Sherpa}

In the \Sherpa event generator~\cite{Gleisberg:2008ta,Bothmann:2019yzt}, photoproduction of jets at \MCatNLO accuracy has been recently implemented and validated against data~\cite{Hoeche:2023gme} and has been published with version 3.0.
Resolved processes can be calculated through interfaces to various different photon PDFs, where the default is the SaS PDF library~\cite{Schuler:1994ft}.

We generated events at Leading Order and \MCatNLO accuracy as described in~\cite{Hoeche:2023gme}. We used \Amegic~\cite{Krauss:2001iv} and \Comix~\cite{Gleisberg:2008fv} for tree-level matrix elements and subtraction terms~\cite{Gleisberg:2007md}, and \OpenLoops~\cite{Buccioni:2019sur} for one-loop matrix elements. For the calculation of the matrix element, we treated the $b$-quarks for \ZEUS, and additionally the $c$-quark for \OPAL and \EIC runs, as massive and included them in the final state. While at LO all subprocesses are available, the initial state subtraction for massive quarks has not been implemented yet.
The \CSS parton shower~\cite{Schumann:2007mg} was used, combined with the \MCatNLO method~\cite{Frixione:2002ik, Frixione:2003ei} as implemented in \Sherpa~\cite{Hoeche:2011fd}.

MPIs were modelled through an implementation of the Sjostrand-van Zijl model~\cite{Sjostrand:1987su,Schuler:1993wr} in which the total hadronic cross-section is calculated using Regge theory and the parametrizations for the photon are obtained as a superposition of light neutral vector mesons as proposed and parametrised in~\cite{Schuler:1993wr}. The tuning of the rewritten MPI modelling is currently work-in-progress as part of the new release. Particles were hadronized through the cluster fragmentation model in \Ahadic~\cite{Chahal:2022rid} and has already been tuned against data.
The photon flux was modelled as computed in~\cite{Frixione:1993yw}, which includes a correction for $x \to 1$ relevant for lepton-hadron colliders. For the PDFs, we used the SAS2M~\cite{Schuler1995,Schuler1996a} set for the photon and the PDF4LHC21\_40\_pdfas~\cite{PDF4LHCWorkingGroup:2022cjn} set for the proton, and the value of $\alpha_{\mathrm{S}}$ was kept at its default value of 0.118 with three-loop running.
Factorisation and renormalisation scale were set to $\mu_{\mathrm{F}} = \mu_{\mathrm{R}} = H_{\mathrm{T}} / 2$ and a 7-point scale variation in the matrix element and the shower was done as an uncertainty estimation.

\section{Differences between generators}\label{sec:diffs}

\begin{figure}[tbhp]
    \centering
    \begin{tabular}{cc}
        \includegraphics[width=0.45\linewidth]{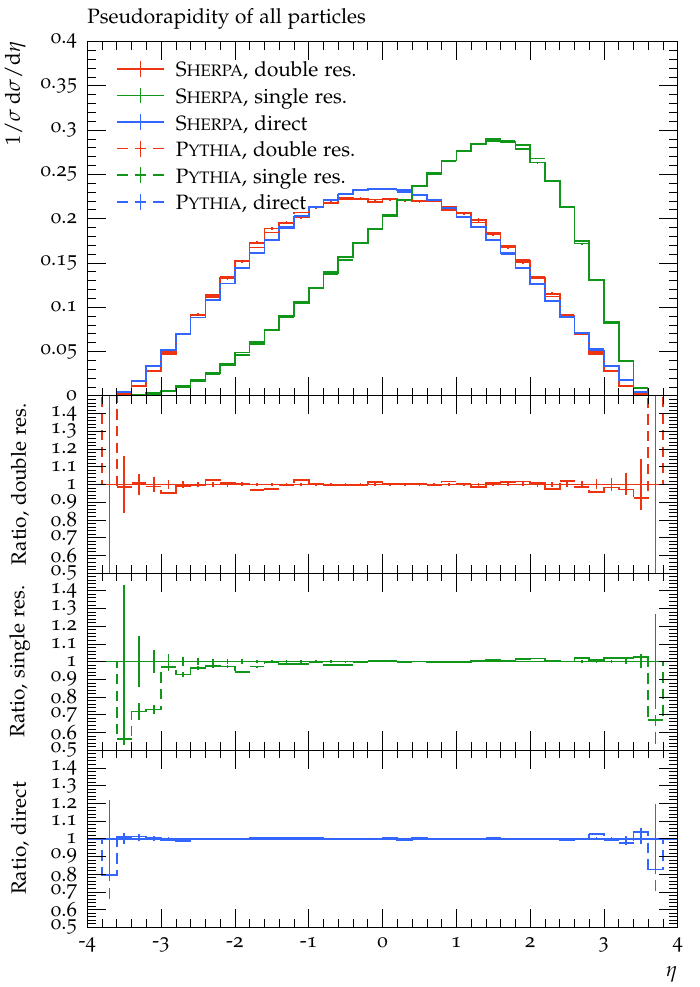} &
        \includegraphics[width=0.45\linewidth]{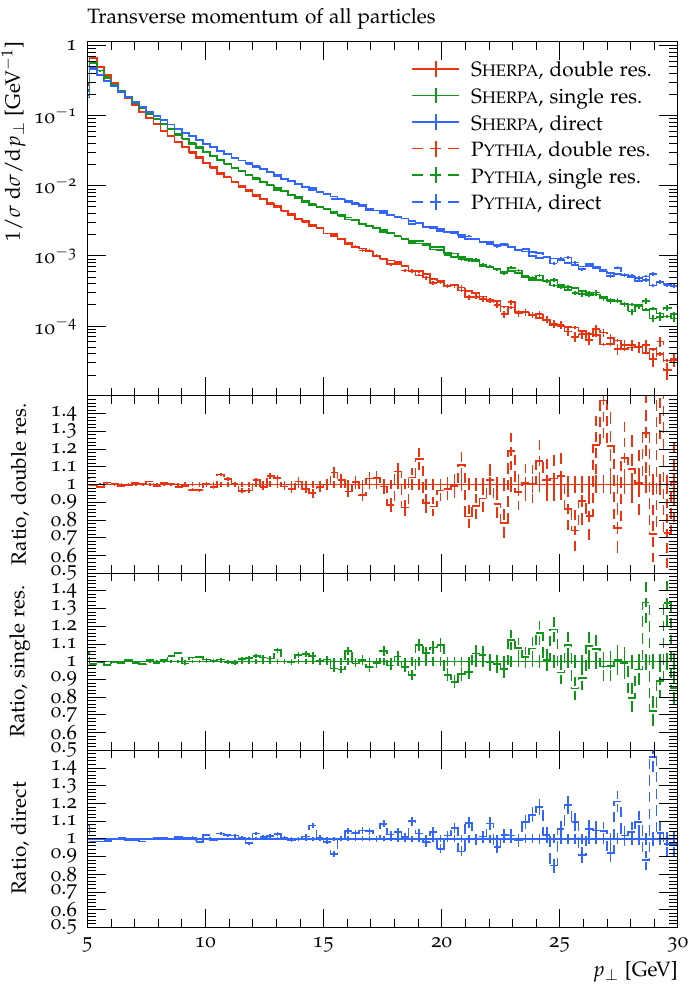}
    \end{tabular}
    \caption{Comparison of fixed-order parton level events between \Pythia and \Sherpa for a \lep-like setup.}\label{fig:bare-partonlevel}
\end{figure}

As a baseline for a broader comparison between the generators, we compared events of \lep-like setups with beam energies of 99 GeV from \Pythia and \Sherpa at the bare cross-section-level for each of the components in Fig.~\ref{fig:bare-partonlevel} (due to technical reasons, \Herwig can not be compared at this level). Here we used the same photon flux, PDF, and a setting of $\alpha_S = 0.130$ with 1-loop running and computed only processes of light partons. The considered observables include the pseudorapidity distribution and transverse momentum spectra of the outgoing partons for direct, single-resolved and double-resolved contributions separately. In case of single-resolved contribution the direct photon has a positive $p_z$. We find that indeed the results are, up to statistical fluctuations, in perfect agreement at this level.

\begin{figure}[hp]
    \centering
        \begin{tabular}{cc}
        \includegraphics[width=0.4\linewidth]{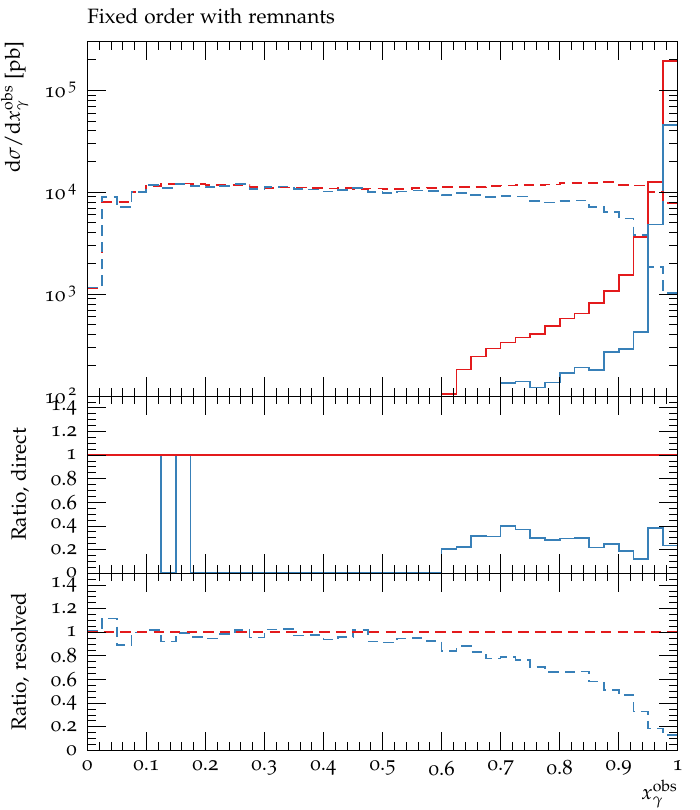} &
        \includegraphics[width=0.4\linewidth]{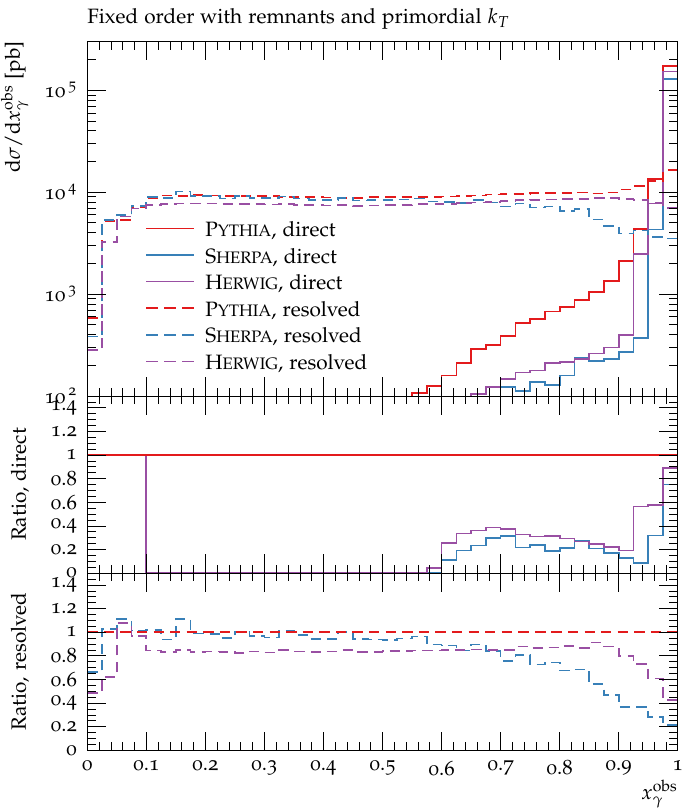} \\ \vspace{-2mm}
        \includegraphics[width=0.4\linewidth]{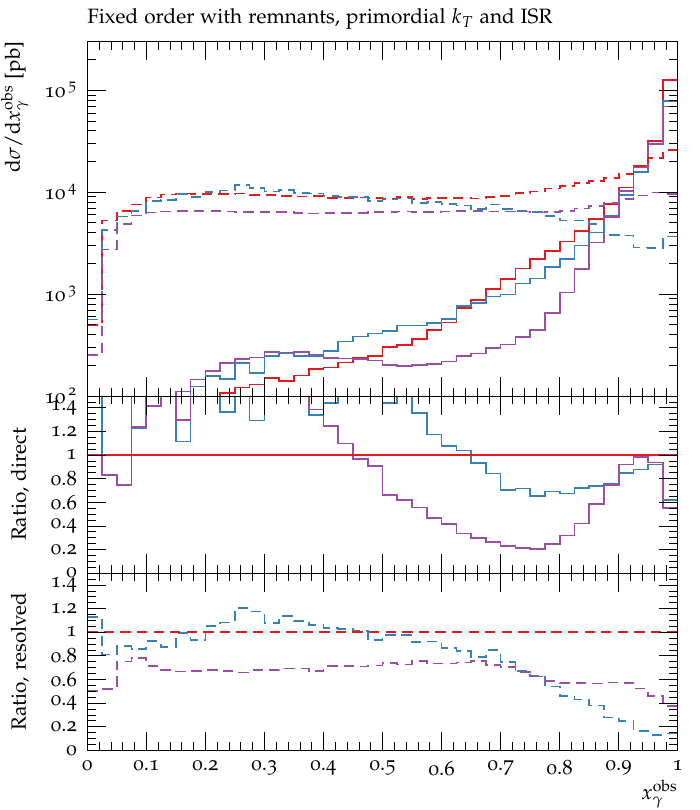} &
        \includegraphics[width=0.4\linewidth]{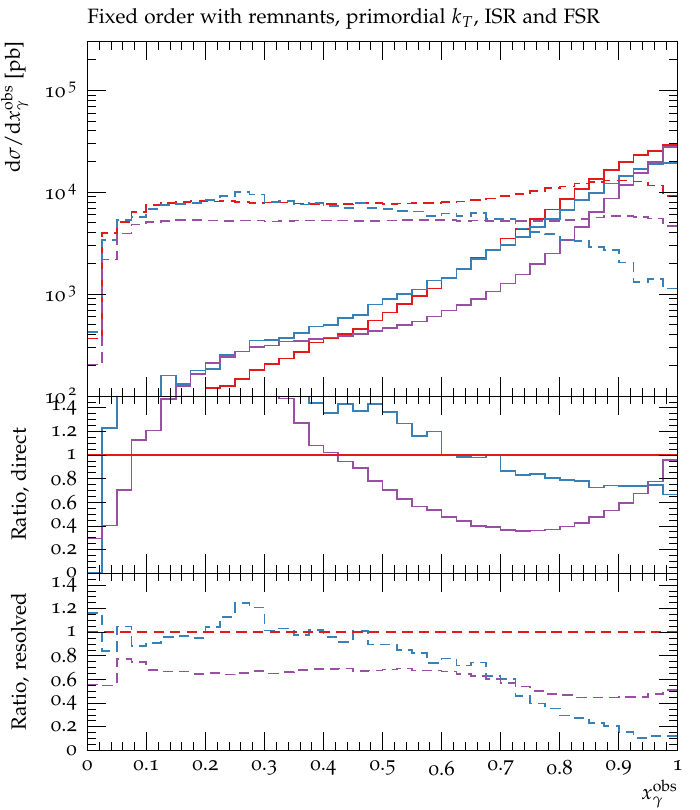} \\ \vspace{-2mm}
        \includegraphics[width=0.4\linewidth]{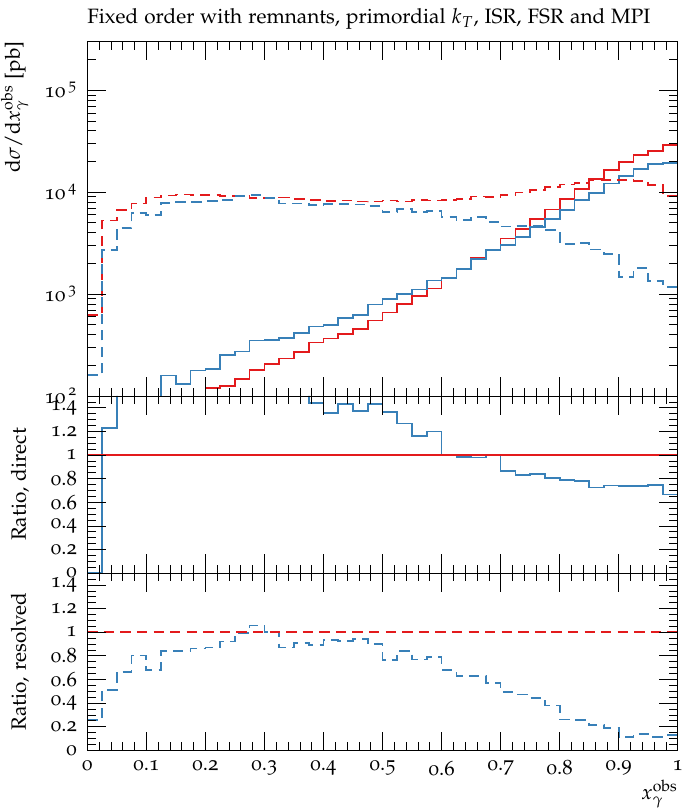} &
         \includegraphics[width=0.4\linewidth]{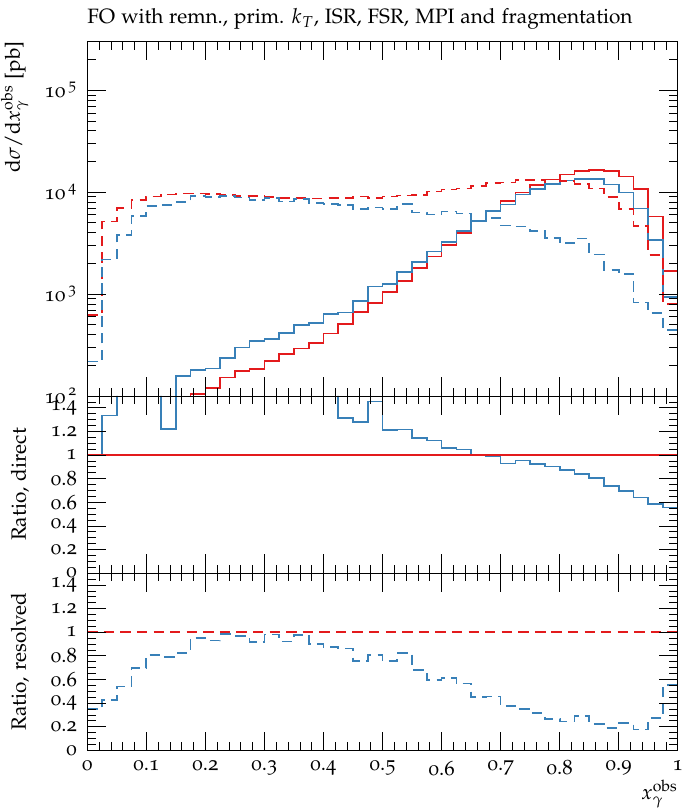}
           \end{tabular}
         \caption{\label{figures:xg-steps}Comparison of the $x_{\gamma}^{\mathrm{obs}}$ distributions at various stages of the event generation between \Pythia, \Sherpa and, if possible, \Herwig. From left to right, top to bottom: fixed order with remnants, plus primordial $k_T$, plus ISR, plus FSR, plus MPI, and finally also including hadronization.}
\end{figure}

In order to quantify the impact of each component in the event generation we continue with comparisons where we include each component one by one. For this we switch from the e$^+$e$^-$ setup of Fig.~\ref{fig:bare-partonlevel} to photoproduction in ep at EIC-like beam energies, $E_{\mathrm{p}} = 275~\text{GeV}$ and $E_{\mathrm{e}} = 18~\text{GeV}$, which is the primary target for photoproduction phenomenology at the \EIC. We apply the same matched settings and non-perturbative inputs and focus on differences in the modelling part.
As the pure hard-process level was already validated with the \LEP setup above, we start
from the hard process with remnants, then successively including intrinsic $k_\perp$, parton-shower for initial and final state radiation (ISR, FSR), and ultimately MPIs and hadronization. We show these comparisons for $x_\gamma^{\mathrm{obs}}$ distributions in Fig.~\ref{figures:xg-steps} where
\begin{equation}
    x_\gamma^\mathrm{obs} = \frac{
        \sum_{j=1,2} E_{\mathrm{T}}^{(j)} \mathrm{e}^{-\eta^{(j)}}}{2 y E_{\mathrm{e}}} \ ,
\end{equation}
and the sum runs over the two highest $E_{\mathrm{T}}$ jets, $E_{\mathrm{e}}$ is the energy of the incoming electron and $y$ the inelasticity.
At LO $2 \rightarrow 2$ kinematics this matches $x_\gamma$, the momentum fraction of partons in a photon, and therefore can also be applied to separate direct and resolved contributions.
It is also ideal observable to highlight a few differences in modelling between different event generators.
The corresponding distributions of transverse momentum and pseudorapidity of all final-state particles at each stage are shown in Figs.~\ref{figures:pT-steps} and~\ref{figures:eta-steps} in the appendix.
In case of \Herwig, it is not possible to leave out certain parts of the modelling in a meaningful manner (due to technical restrictions on the event record handling in particular regarding remnants) and thus we only show these comparisons for a subset of the event generation stages.

When adding the remnant partons in addition to the hard process, c.f.\ first panel in Fig.~\ref{figures:xg-steps}, we notice that this will cut some amount of cross section from the resolved processes at high-$x_{\gamma}^{\mathrm{obs}}$ end. In \Pythia the effect is fairly modest but more pronounced in \Sherpa. Including primordial $k_{\perp}$ for the remnants and hard-process initiators brings the large-$x_{\gamma}^{\mathrm{obs}}$ cross section up a bit where now \Herwig sits between \Pythia and \Sherpa. Adding ISR shifts the cross section of the resolved channel towards larger $x_{\gamma}^{\mathrm{obs}}$ in \Pythia but not for \Sherpa or \Herwig, which can be attributed to the $\gamma \rightarrow q\bar{q}$ term in the ISR as will be discussed below. Parton showers also make the direct contributions wider and FSR translate events from higher to lower values of $x_{\gamma}^{\mathrm{obs}}$. MPIs (not enabled in \Herwig) increase the cross section at low-$x_{\gamma}^{\mathrm{obs}}$ region simply since for these events there are more energy remaining for subsequent partonic interactions after the primary one and we find the effect slightly more pronounced in \Pythia compared to \Sherpa. Hadronization will further push events towards lower $x_{\gamma}^{\mathrm{obs}}$ but the separation between \Pythia and \Sherpa for the resolved events generated by the remnants and ISR implementation survives to hadron level whereas \Herwig would tends to sit in between the others at $x_{\gamma}^{\mathrm{obs}}>0.7$ and slightly below the two at low $x_{\gamma}^{\mathrm{obs}}<0.5$.
Hence we generally see that this observable, while particularly useful for photon PDF fits due to its easy interpretation, is also sensitive to logarithmic corrections from the parton shower and non-perturbative parameters like primordial $k_{\perp}$, hadronisation, and MPI.
\begin{figure}[hp]
    \centering
        \begin{tabular}{cc}
        \includegraphics[width=0.45\linewidth]{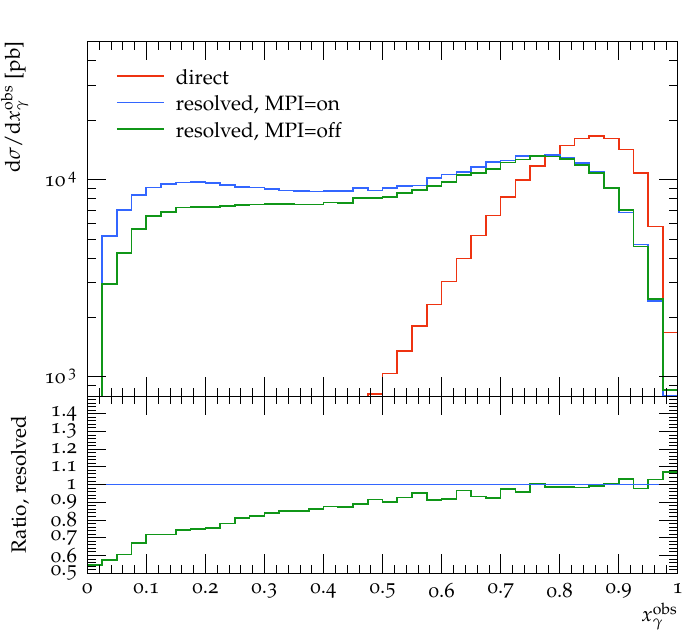} &
        \includegraphics[width=0.45\linewidth]{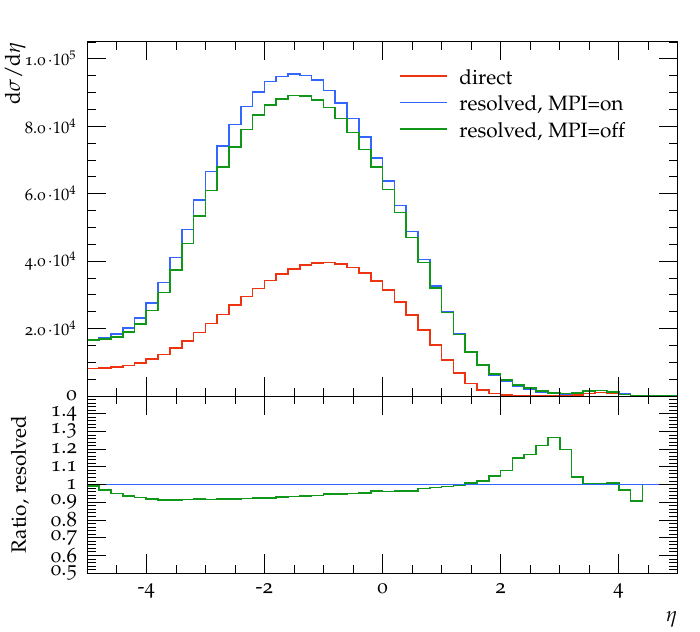} \\
        \includegraphics[width=0.45\linewidth]{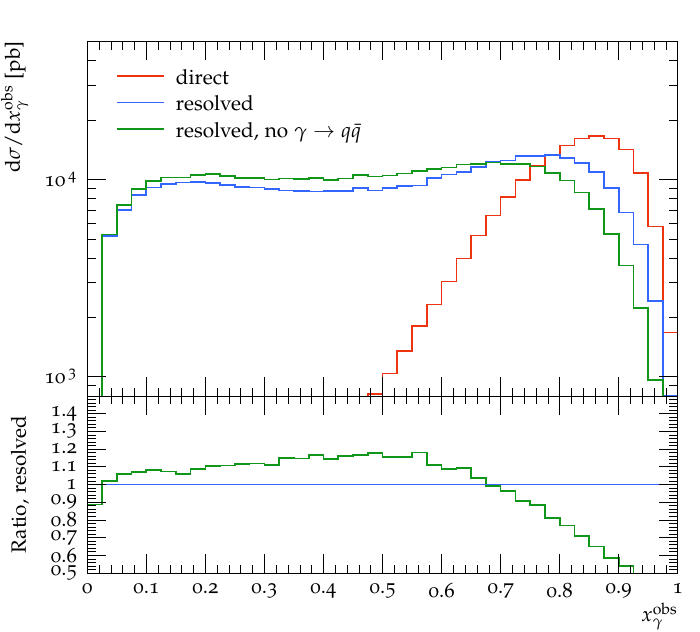} &
        \includegraphics[width=0.45\linewidth]{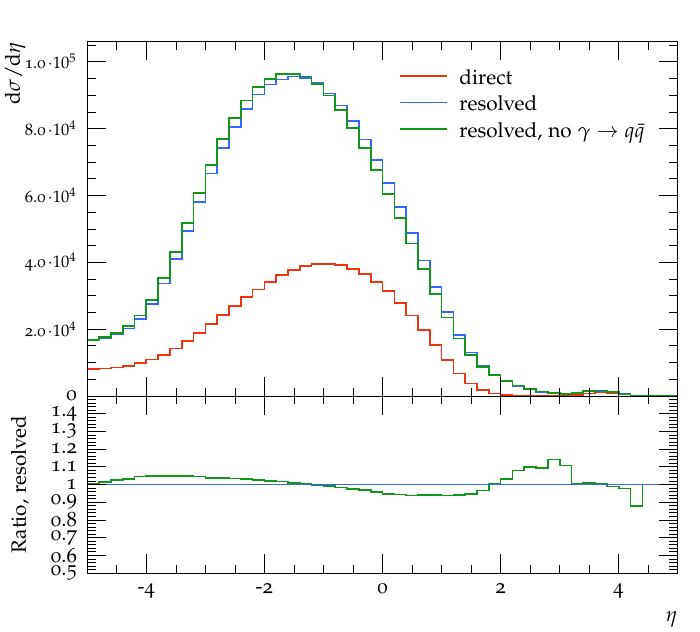} \\
        \includegraphics[width=0.45\linewidth]{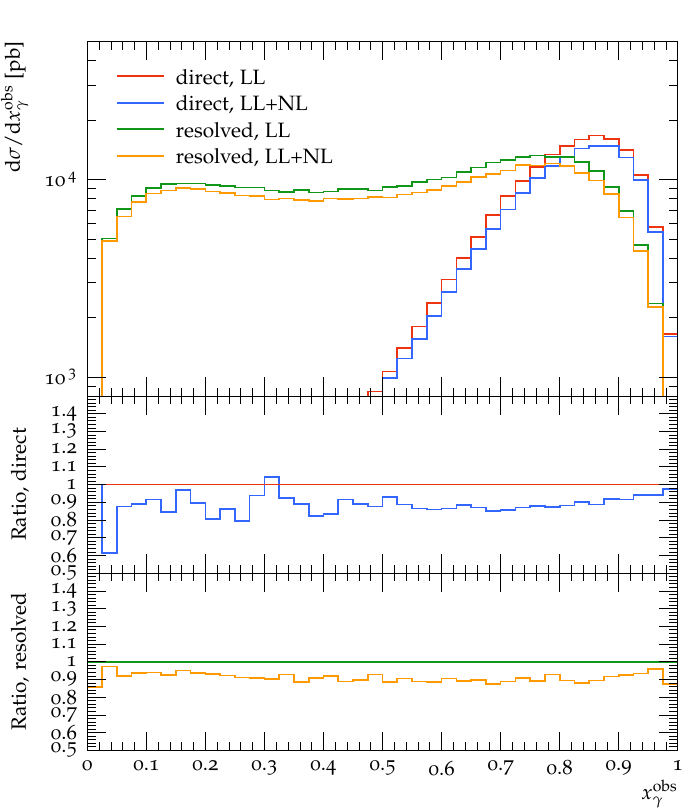} &
        \includegraphics[width=0.45\linewidth]{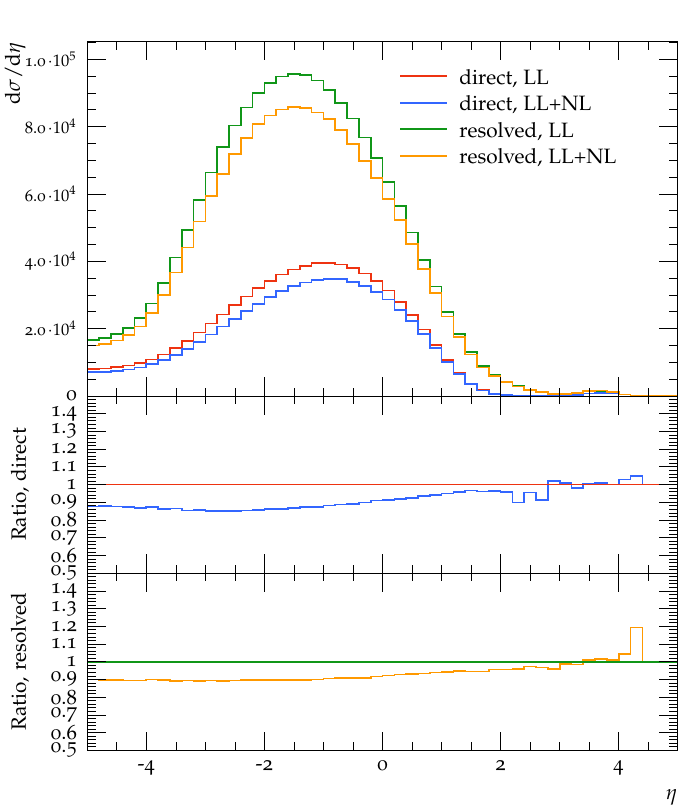}
           \end{tabular}
         \caption{\label{figures:py8-diag}Pythia simulations without MPIs (top), without $\gamma \rightarrow q\bar{q}$ term in the ISR (middle) and with non-logarithmic term in the photon flux (bottom) compared to default settings for the full simulations with the EIC kinematics for $x_{\gamma}^{\mathrm{obs}}$ (left) and $\eta$ distribution of all particles (right).}
\end{figure}

In order to quantify effects from MPIs, $\gamma \rightarrow q\bar{q}$ splitting in the ISR algorithm and the NL correction to the photon flux, we present comparison from \Pythia with and without these effects in Fig.~\ref{figures:py8-diag}. We consider here $x_{\gamma}^{\mathrm{obs}}$ and $\eta$ distributions for full hadron-level simulations with the same EIC-like setup as above. Enabling MPIs increases jet-production cross section for resolved events at small-$x_{\gamma}^{\mathrm{obs}}$ around 50\% but the impact for the multiplicity is fairly modest, less than 10\%.
The inclusion of the $\gamma \rightarrow q\bar{q}$ splitting in the ISR, which is currently only available in the \Pythia default parton shower\footnote{This splitting is always enabled in \Pythia and cannot be disabled without modifying the source code.}, shifts events to higher $x_{\gamma}^{\mathrm{obs}}$. In particular, it leads to a significant increase of the cross section at $x_{\gamma}^{\mathrm{obs}} > 0.8$ for the resolved contribution. Similarly $\eta$ distribution is shifted towards proton-going direction but the effect is modest.
Adding NL correction to the photon flux will reduce jet cross sections by up to 10\% fairly equally over the whole $x_{\gamma}^{\mathrm{obs}}$ kinematics for both resolved and direct contributions. The inclusive hadron production is reduced by a similar fraction and the reduction is more pronounced at negative pseudorapidities. Since \Sherpa already applies an equivalent flux with NL corrections by default (Sec.~\ref{sec:MCs}), this reduction is built into the \Sherpa predictions and contributes to the systematic normalisation offset between \Pythia and \Sherpa seen in the comparisons below.

Even though not specific to photoproduction we point out that the settings of $\alpha_{S}$ are different: while in \Sherpa the current default is set according to the default proton PDF set and close to the PDG world average~\cite{ParticleDataGroup:2020ssz}, in \Pythia it is being used as a dynamical $K$ factor, also in accordance with its default proton PDF set. Photon PDFs do not allow the nowadays standard procedure of setting $\alpha_{S}$ in accordance with the fit as this information is often not given in the corresponding publication and would in any case be in conflict with settings of modern proton PDFs. Hence, strictly speaking the factorisation of the cross section is not fully consistent due to the different values of $\alpha_{S}$ used throughout event generation.

\begin{figure}[ht]
    \centering
    \includegraphics[width=.6\linewidth]{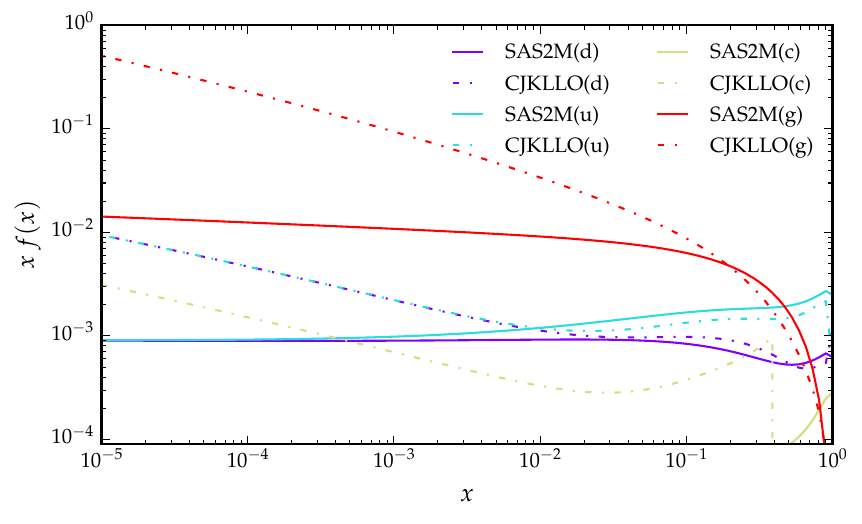}
    \caption{Comparison of the CJKLLO and SAS2M PDF sets for $u$, $d$ and $c$ quarks and the gluon at $\mu_{\mathrm{F}}^2= 5\ \mathrm{GeV}^2$. }\label{fig:pdf-plot}
\end{figure}

As a last point, we plot the parton distributions of the light partons for the two PDF libraries SAS2M and CJKL in Fig.~\ref{fig:pdf-plot}. The two fits come to vastly different behaviour at small $x_{\gamma}$, where the SAS2M sets reach an almost constant value and the CJKL sets behave like a power-law, with the most pronounced difference in the gluon distribution. Since the gluon dominates the resolved photon cross section at low $E_{\mathrm{T}}$ and forward rapidities, this difference is expected to contribute to the cross section difference between \Pythia, \Herwig and \Sherpa in the data comparisons below.

In Tab.~\ref{tab:differences}, we summarise the differences between the generators with respect to the simulation of photoproduction. Other differences are for example the fragmentation model and QED corrections, however, these will not be discussed here and can be found in the manuals~\cite{Bahr:2008pv,sherpa-manual,Bierlich:2022pfr}. It is also worth to point out that unlike the settings for $\alpha_S$ and the flux, which cover the perturbative region and are independent of any fitting, the remnant creation and the MPIs interplay with non-perturbative effects and rely on precise data by experiments like \hera{} and \lep{}.

\begin{table}[htbp]
    \centering
    \begin{tabular}{|l|l|l|l|} \hline
        Property & \Pythia & \Sherpa & \Herwig \\ \hline\hline
        Flux & LL & LL+NL & LL \\
        $\alpha_S(M_{\mathrm{Z}}^2)$ & 0.130, 1-loop running & 0.118, 3-loop running & 0.118, 2-loop running \\
        PDFs & CJKL & SAS2M & SAS2M \\
        Remnants & forced splittings/PS rejection & PS rejection & forced splitting \\
        Photon Splitting & yes & no & no\\
        MPI tuning & preliminary $\gamma \mathrm{p}$/$\gamma \gamma$ tune & untuned & untuned \\ \hline
    \end{tabular}
    \caption{Differences between the generators of default settings, specifically for photoproduction.}\label{tab:differences}
\end{table}

\section{Comparisons to \protect\LEP}\label{sec:lep}

In the following sections, each generator is run with its default photoproduction configuration.
Differences in the comparisons therefore reflect the combined effect of individual models \textit{and} varying inputs, c.f. Tab.~\ref{tab:differences}.

For the comparison to \LEP data, we used a dijet measurement from the \OPAL collaboration~\cite{OPAL:2003hoh}. At lepton colliders, the cross-section for photoproduction can be separated into $\sigma = \sigma_{\gamma \gamma \to X} + \sigma_{\gamma j \to X} + \sigma_{j \gamma \to X} + \sigma_{j j \to X}$, where $j$ denotes a photon being resolved into partons. The analysis used data taken at $\sqrt{s} = 198$ GeV and clustered jets with the $k_{\mathrm{T}}$ algorithm with $R = 1$ demanding $E_{\mathrm{T}} > 3$ GeV and $|\eta | < 2$ with at least two jets. To separate resolved from direct photoproduction processes, the analysis defined
\begin{equation}
    x_\gamma^\pm = \frac{
            \sum_{j=1,2} E^{(j)} \pm p_z^{(j)}
        }{
            \sum_{i \in \mathrm{hfs}} E^{(i)} \pm p_z^{(i)}
        }
\end{equation}
and associated values $x_\gamma^\pm < 0.75$ with resolved processes. For a parton-level $2 \rightarrow 2$ scattering these definitions would again match with the momentum fractions in photons going positive and negative $p_z$ in Eq.~\ref{eq:sigmahardEE} but adding parton showers, MPIs and hadronization will smear the kinematics such that the division will be only approximative. In the numerator the sum runs over the two jets with the highest transverse momentum, $p_{\mathrm{T}}$, and in the denominator over the complete hadronic final state.

\begin{figure}
    \centering
    \begin{tabular}{ccc}
        \includegraphics[width=0.3\linewidth]{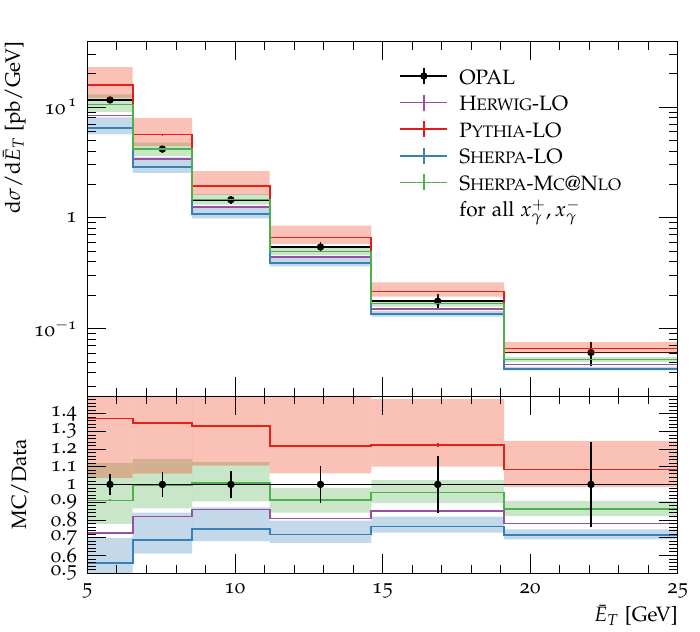} &
        \includegraphics[width=0.3\linewidth]{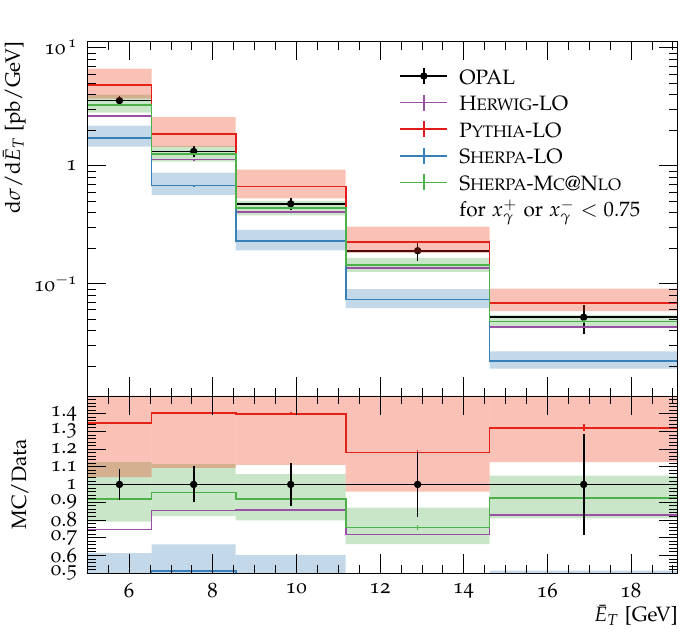} &
        \includegraphics[width=0.3\linewidth]{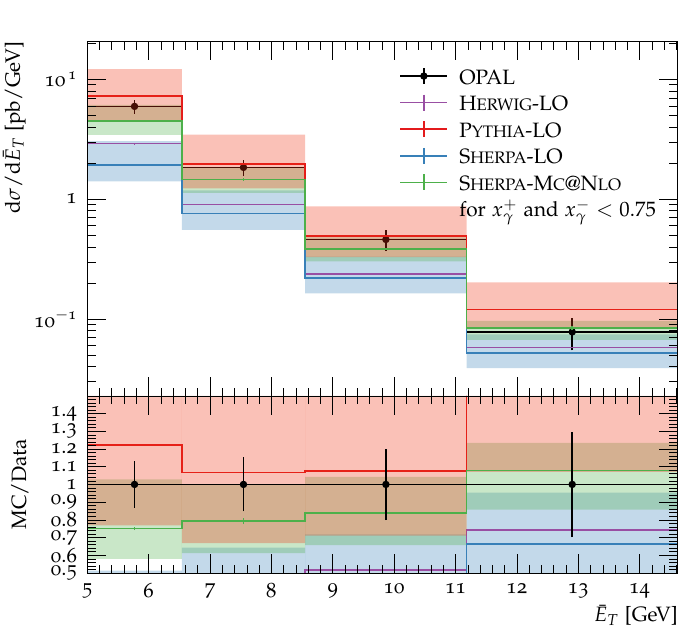}
    \end{tabular}
    \caption{Distribution of average transverse energy of di-jets, $\bar E_T$, from \protect\OPAL~\protect\cite{OPAL:2003hoh} for all $x^\pm_\gamma$ (left), $x^+_\gamma$ or $x^-_\gamma < 0.75$ (middle) and $x^\pm_\gamma < 0.75$ (right), compared to Leading Order simulations by \protect\Herwig{}, \protect\Pythia and \protect\Sherpa and \protect\MCatNLO-accurate simulations by \protect\Sherpa.}\label{fig:opal-et}
\end{figure}

In Fig.~\ref{fig:opal-et} we present the average transverse energy of the di-jets for different taggings on $x_\gamma^\pm$. The LO prediction from \Sherpa{} undershoots the total cross-section, however, the effect is more pronounced for resolved photons than for unresolved ones.
Going to NLO accuracy, the simulation describes the data well within the errors and the scale uncertainties are significantly reduced by about a factor 2. The large $K$ factors hint at the real corrections and the filled-up phase space as the drivers of the improved description. \Pythia{} tends to overshoot the data for events with direct contribution but agrees well with the data for the resolved-resolved case. The \Herwig simulation (without MPI for resolved photons, cf.~Sec.~\ref{sec:MCs}) typically lies between the \Sherpa and \Pythia results. In all cases the data is within the large uncertainties from scale variations. As discussed in Sec.~\ref{sec:diffs}, the large difference between the LO results from \Herwig, \Pythia and \Sherpa builds up from varying inputs.

\begin{figure}
    \centering
    \begin{tabular}{cc}
        \includegraphics[width=0.4\linewidth]{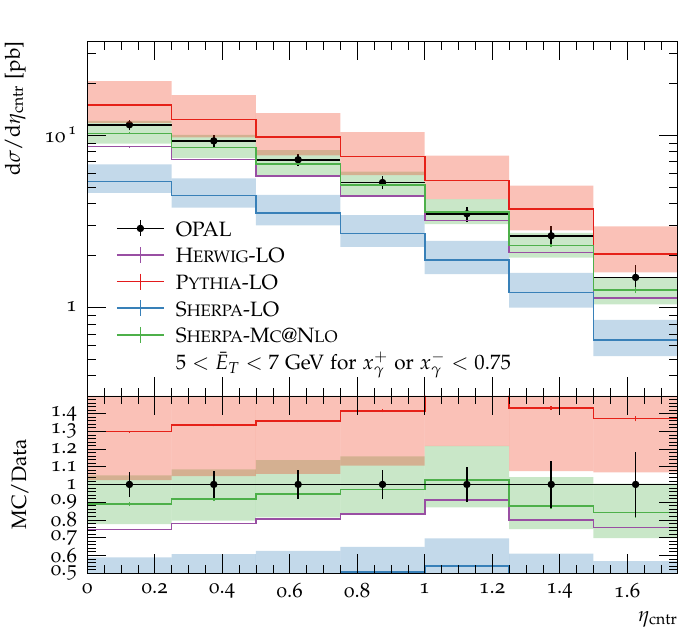} &
        \includegraphics[width=0.4\linewidth]{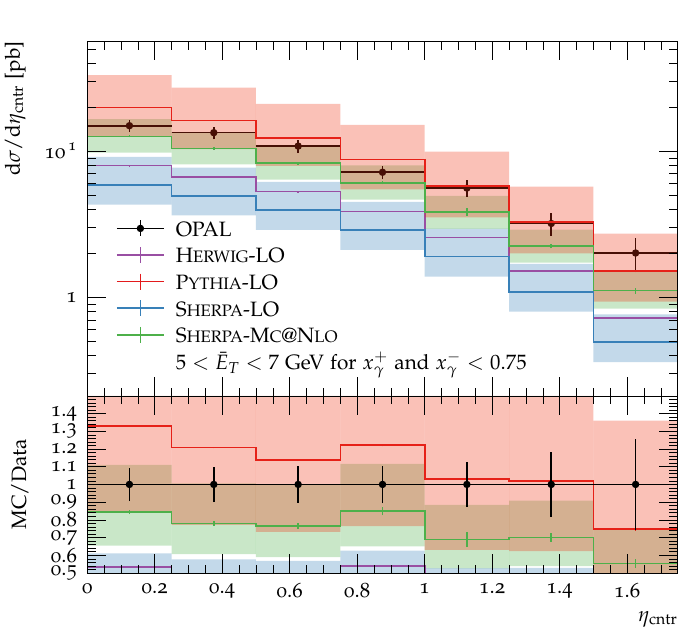}
    \end{tabular}
    \caption{Distribution of pseudo-rapidity $\eta$ of di-jets from \protect\OPAL~\protect\cite{OPAL:2003hoh} for $x^+_\gamma$ or $x^-_\gamma < 0.75$ (left) and $x^\pm_\gamma < 0.75$ (right), compared to Leading Order simulations by \protect\Herwig, \protect\Pythia and \protect\Sherpa and \protect\MCatNLO-accurate simulations by \protect\Sherpa.}\label{fig:opal-eta}
\end{figure}

The distributions in pseudo-rapidity $\eta$ in Fig.~\ref{fig:opal-eta} show a similar picture, however for double-resolved processes, \ie $x^\pm_\gamma < 0.75$, all predictions
have a slightly steeper fall-off as a function of $\eta$ than seen in the data,
leaning towards an undershoot in the forward region. One potential reason could be the weakly constrained gluon-content of the photon, which should be the leading contribution at low-$E_{\mathrm{T}}$ and forward jets.

\begin{figure}
    \centering
    \begin{tabular}{cc}
        \includegraphics[width=0.4\linewidth]{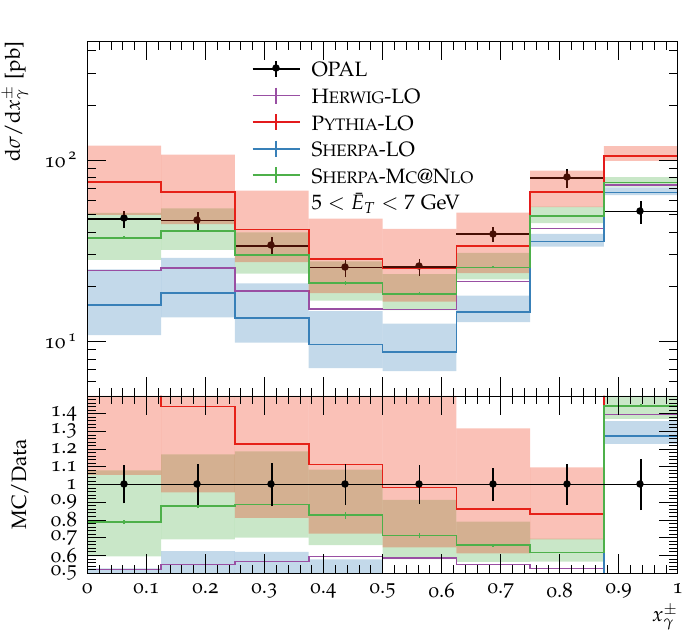} &
        \includegraphics[width=0.4\linewidth]{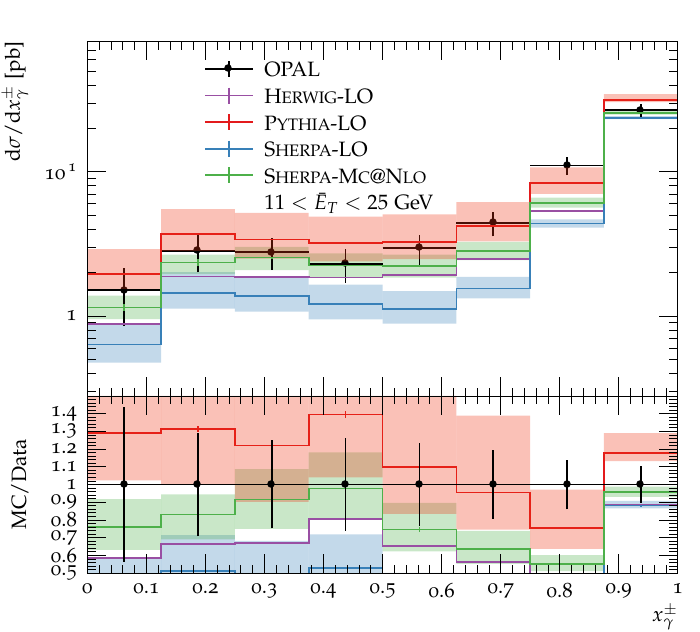}
    \end{tabular}
    \caption{Distribution of $x_\gamma^{\pm}$ of di-jets from \protect\OPAL~\protect\cite{OPAL:2003hoh} for bins of average jet transverse energy $\bar E_{\mathrm{T}}\in [5\,{\rm GeV}, 7\,{\rm GeV}]$ (left) and $\bar E_{\mathrm{T}}\in [11\,{\rm GeV}, 25\,{\rm GeV}]$ (right), compared to Leading Order simulations by \protect\Herwig{}, \protect\Pythia and \protect\Sherpa and \protect\MCatNLO-accurate simulations by \protect\Sherpa.}\label{fig:opal-xgamma}
\end{figure}

Fig.~\ref{fig:opal-xgamma} shows distributions of $x_\gamma^{\pm}$ for low- and high-average jet transverse energies $\bar E_{\mathrm{T}}$, respectively. We see good agreement for both the \Pythia and the \Sherpa-\MCatNLO simulations; however, the transition from the resolved to the direct processes seems to be poorly modelled, as can be seen by the consistent undershoot at around $x_\gamma^\pm \approx 0.8$ for all predictions. Unlike \Herwig and \Sherpa, the \Pythia simulation does include the correct evolution of the photon PDF, \ie{} the photon splitting $\gamma \to q \bar q$ is taken into account, the distribution still shows this shape in that case too, though. Inversely, the NLO correction in \Sherpa does not fill this dip either, hence it is not a feature due to lacking higher orders or unfilled phase space. Potential reasons could be the poorly constrained photon PDFs or insufficient tuning of the fragmentation modelling. Combined with the overshoot in the largest-$x_{\gamma}^\pm$ bin, we would expect these effects to increase the multiplicity of direct processes, hence shifting cross-section towards lower values of $x_{\gamma}^\pm$. It would be informative to first ensure that it is not an artifact of poorly constrained parameters in the PDF fit by doing a study with a future modern PDF parametrisation, before analysing the dependence on, e.g., hadronisation fits or other non-perturbative effects.

\section{Comparisons to \protect\HERA}\label{sec:hera}

For this comparison we used data taken by the \ZEUS collaboration~\cite{ZEUS:2001zoq} during \HERA Run 1. Here, the photoproduction cross-section can be decomposed into two parts, $\sigma = \sigma_{\gamma P \to X} + \sigma_{j P \to X}$, where again $j$ denotes a parton resolved from within the photon. Similar to the previous analysis, jets were clustered with the $k_{\mathrm{T}}$ algorithm with $R=1$ with cuts of $\eta \in [-1, 2.4]$ and $E_{\mathrm{T}} > 14 \ (11)$ GeV for the (sub-)leading jet, respectively.

\begin{figure}
    \centering
    \begin{tabular}{cc}
        \includegraphics[width=0.4\linewidth]{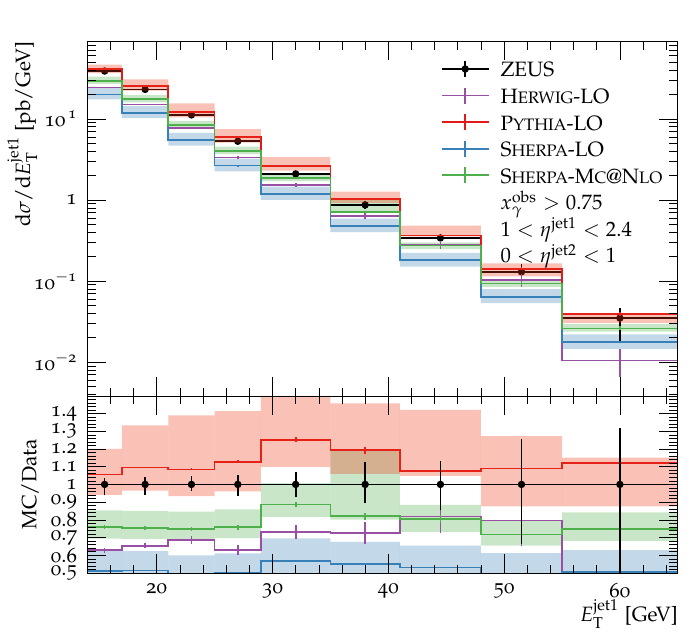} &
        \includegraphics[width=0.4\linewidth]{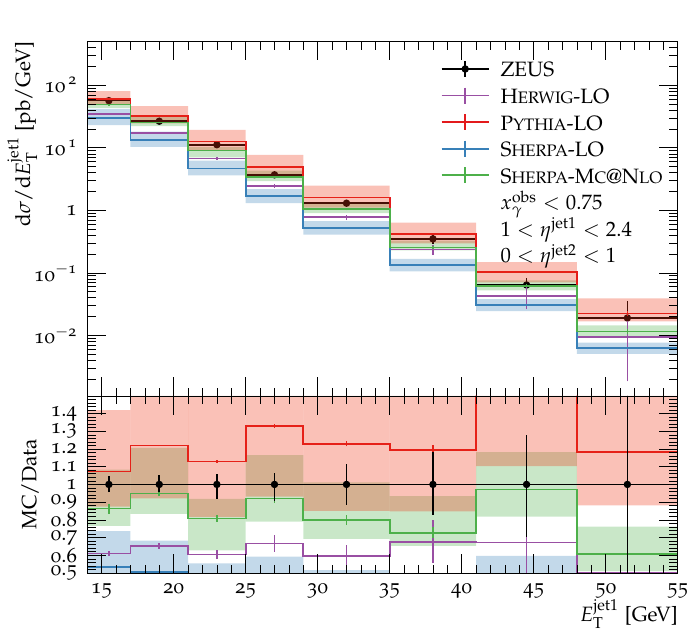}
    \end{tabular}
    \caption{Distributions of leading jet transverse energy $E_{\mathrm{T}}^\mathrm{jet1}$ for $x_\gamma^{\mathrm{obs}} > 0.75$ (left) and $x_\gamma^{\mathrm{obs}} < 0.75$ (right) with jet pseudo-rapidities for the two leading jets in $1 < \eta^\mathrm{jet1} < 2.4$ and $0 < \eta^\mathrm{jet2} < 1$ from \protect\ZEUS~\protect\cite{ZEUS:2001zoq}, compared to Leading Order simulations by \protect\Herwig, \protect\Pythia and \protect\Sherpa and \protect\MCatNLO-accurate simulations by \protect\Sherpa.}\label{fig:zeus-et}
\end{figure}

\begin{figure}
    \centering
    \begin{tabular}{cc}
        \includegraphics[width=0.4\linewidth]{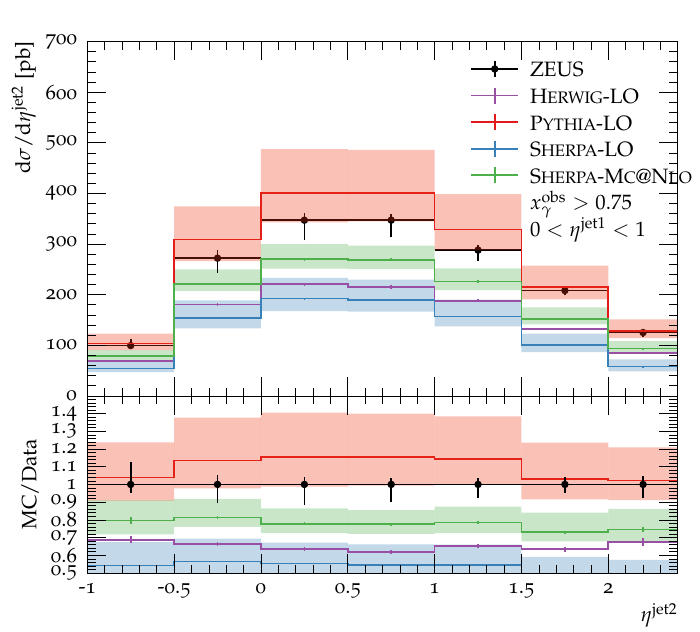} &
        \includegraphics[width=0.4\linewidth]{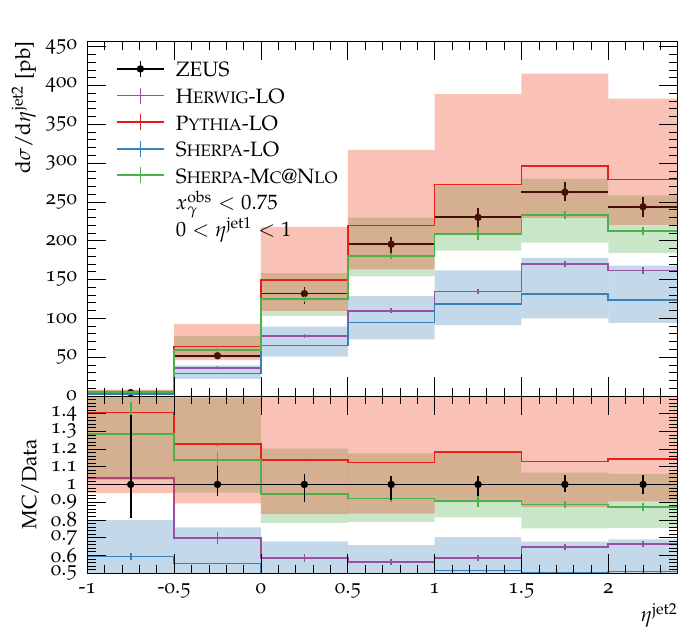}
    \end{tabular}
    \caption{Distributions of sub-leading jet pseudo-rapidity $\eta^\mathrm{jet2}$ for $x_\gamma^{\mathrm{obs}} > 0.75$ (left) and $x_\gamma^{\mathrm{obs}} < 0.75$ (right) with leading jet pseudo-rapidity in $0 < \eta^\mathrm{jet1} < 1$ from \protect\ZEUS~\protect\cite{ZEUS:2001zoq}, compared to Leading Order simulations by \protect\Herwig, \protect\Pythia and \protect\Sherpa and \protect\MCatNLO-accurate simulations by \protect\Sherpa.}\label{fig:zeus-eta}
\end{figure}

In Fig.~\ref{fig:zeus-et} we show the leading jet transverse energy $E_{\mathrm{T}}^\mathrm{jet1}$ for direct and resolved processes. While the \Sherpa-LO prediction stays below the data, the \Pythia predictions agree within the error bars; the \Sherpa-\MCatNLO predictions have again significantly reduced scale uncertainties and describe the data well with the exception of the low-$E_{\mathrm{T}}$ phase space in the direct process, where we see an undershoot of about 20\%. As the cuts on the pseudo-rapidity select the forward region, this observable is probably sensitive to additional radiation from underlying events, the used PDFs, and to the photon splitting in the parton shower. Fig.~\ref{fig:zeus-eta} shows a similar situation for the $\eta$-dependence, where for the resolved processes both, \Pythia and \Sherpa-\MCatNLO, agree with data but for direct processes we see an undershoot with \Sherpa.

\begin{figure}
    \centering
    \begin{tabular}{cc}
        \includegraphics[width=0.4\linewidth]{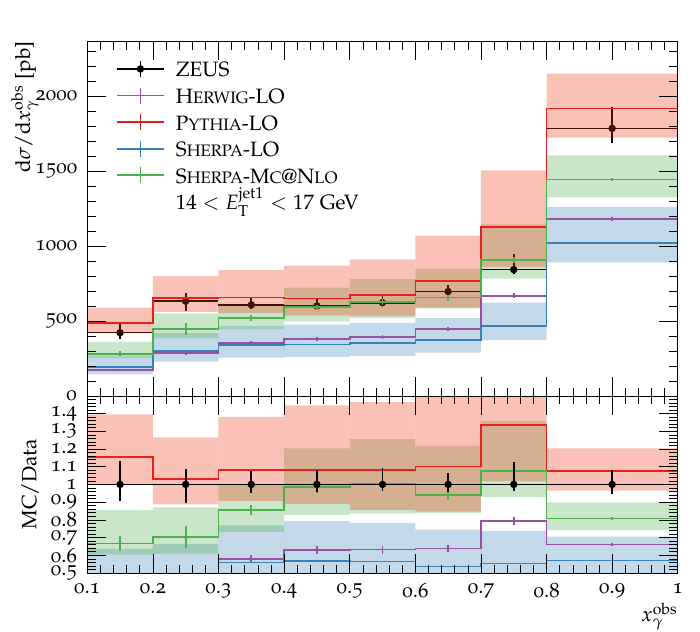} &
        \includegraphics[width=0.4\linewidth]{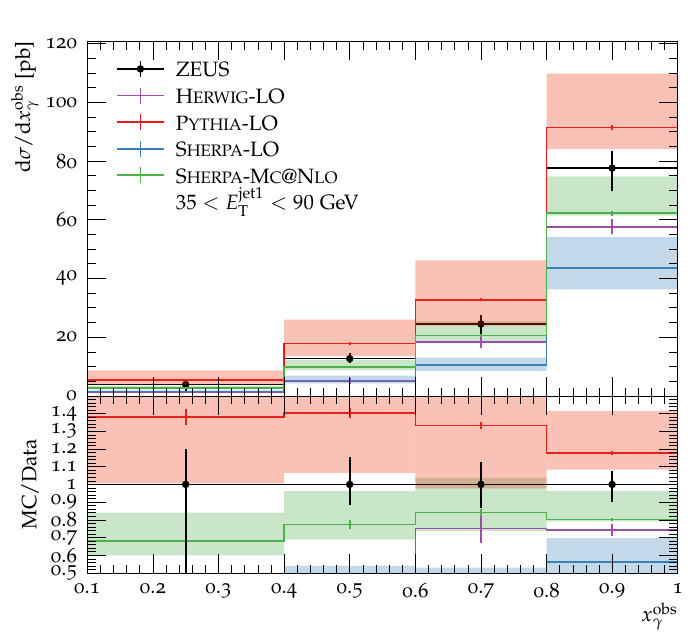}
    \end{tabular}
    \caption{Distributions of $x_\gamma^{\mathrm{obs}}$ for leading jet transverse energies in $14 < E_{\mathrm{T}}^\mathrm{jet1} < 17$ GeV (left) and $35 < E_{\mathrm{T}}^\mathrm{jet1} < 90$ GeV (right) from \protect\ZEUS~\protect\cite{ZEUS:2001zoq}, compared to Leading Order simulations by \protect\Herwig, \protect\Pythia and \protect\Sherpa and \protect\MCatNLO-accurate simulations by \protect\Sherpa.}\label{fig:zeus-xgamma}
\end{figure}

We finish this section with the distributions in $x_\gamma^{\mathrm{obs}}$ for low- and high-leading jet transverse energy in Fig.~\ref{fig:zeus-xgamma}. Opposed to the modelling for \LEP, there is no undershoot visible at the transition from direct to resolved processes. \Sherpa undershoots the data in the region where both $x_\gamma^{\mathrm{obs}}$ and $E_{\mathrm{T}}$ are small, however this can be attributed to the missing tuning of the MPIs as the same region is fairly well described by \Pythia. \Herwig, which lacks MPI for resolved photons (Sec.~\ref{sec:MCs}), is again compatible with the leading order simulation from \Sherpa.

\section{Predictions for \protect\EIC}\label{sec:eic}

\begin{figure}
    \centering
    \begin{tabular}{cc}
        \includegraphics[width=0.4\linewidth]{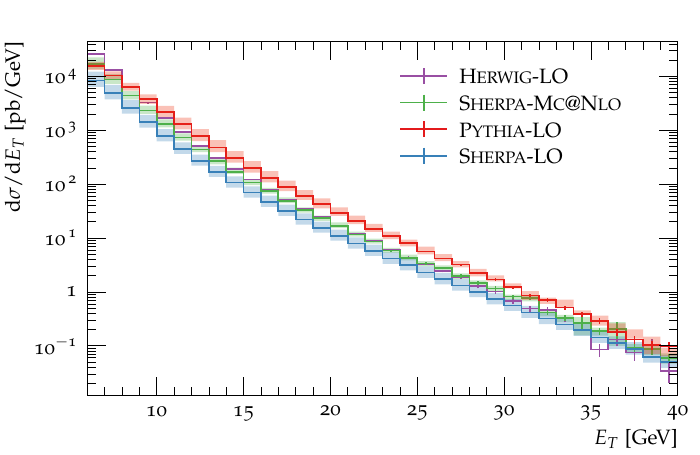} &
        \includegraphics[width=0.4\linewidth]{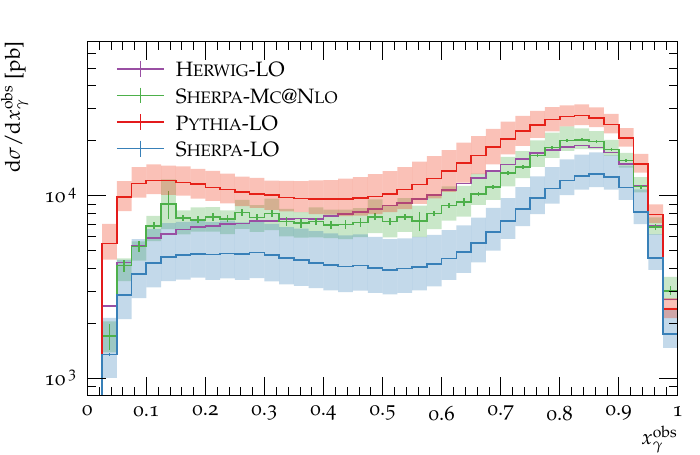}
    \end{tabular}
    \caption{Predictions of transverse jet energy $E_T$ (left) and $x_\gamma^{\mathrm{obs}}$ (right) for jet production at the \EIC, comparing to Leading Order simulations by \protect\Herwig, \protect\Pythia and \protect\Sherpa and \protect\MCatNLO-accurate simulations by \protect\Sherpa.}\label{fig:eic-xgamma}
\end{figure}

For the planned \EIC we present predictions for electron-proton beams with 18 and 275 GeV beam energies, respectively, similar to the study in~\cite{Meinzinger:2023xuf}. We cluster jets with the anti-$k_{\mathrm{T}}$ algorithm with $R = 1.0$ and demand at least one jet with $E_{\mathrm{T}} > 6$ GeV.
Looking at inclusive (di-)jet observables in Fig.~\ref{fig:eic-xgamma}, we see a similar behaviour in the comparison between the generators, where \Sherpa-LO yields the smallest cross-section and \Pythia-LO the largest, while \Herwig delivers a significantly different shape of the $x_\gamma^{\mathrm{obs}}$ distribution. The $K$ factor in these observables is roughly 50\%, again hinting at the real correction and the phase space driving the correction at NLO. The \Pythia LO prediction deviates another 50\% from the NLO-accurate prediction, which can partially be explained by the different PDF sets, $\alpha_{S}$ value and the other differences as pointed out in Sec.~\ref{sec:diffs}. This means that, going towards highest possible precision, the perturbative accuracy needs to be improved further and the non-perturbative effects need to be constrained by data.
Publicly available \Rivet analyses for the relevant data from \hera and \lep are crucial to reach the latter goal. While there has been some recent progress on porting the existing analyses to \Rivet framework, there still are some shortages related to MPI constraints and virtuality modelling.

\begin{figure}
    \centering
    \begin{tabular}{cc}
        \includegraphics[width=0.4\linewidth]{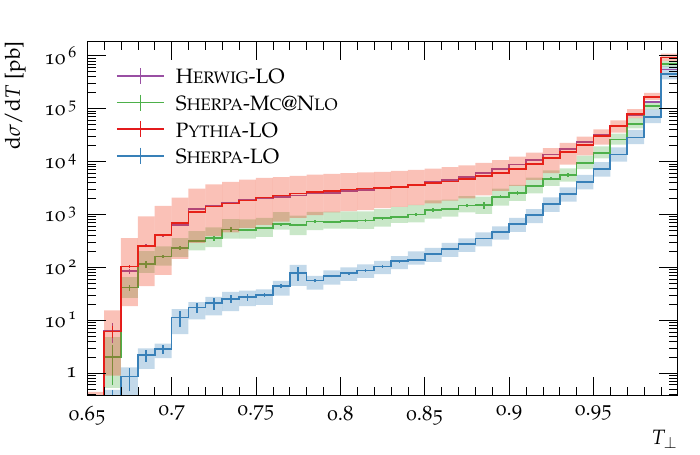} &
        \includegraphics[width=0.4\linewidth]{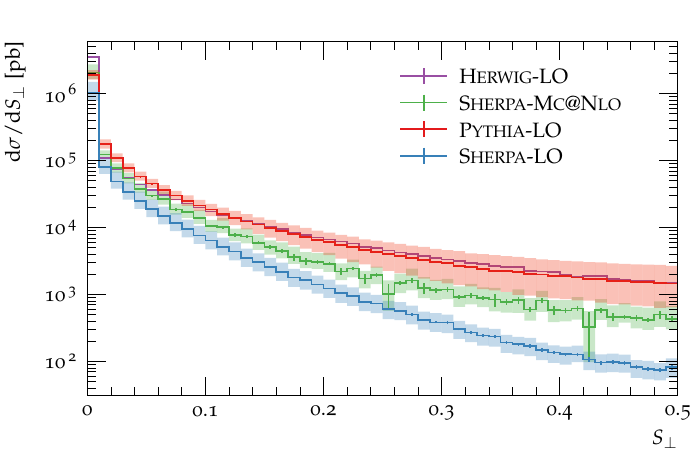}
    \end{tabular}
    \caption{Predictions of transverse thrust $T_\perp$ (left) and transverse sphericity $S_\perp$ (right), for jet production at the \EIC, comparing to Leading Order simulations by \protect\Herwig, \protect\Pythia and \protect\Sherpa and \protect\MCatNLO-accurate simulations by \protect\Sherpa.}\label{fig:eic-thrust}
\end{figure}

To study the event shapes in more detail, we present predictions for transverse thrust $T_\perp$ and transverse sphericity $S_\perp$, defined as
\begin{align}
T_\perp &= \mathrm{max}_{\vec n_{\mathrm{T}}} \frac{\sum_i \left| \vec p_{\mathrm{T},i} \cdot \vec n_{\mathrm{T}} \right|}{\sum_i \left| \vec p_{\mathrm{T},i} \right|} \\
S_\perp &= \frac{2 \lambda_2}{\lambda_1 + \lambda_2}
\end{align}
in Fig.~\ref{fig:eic-thrust}. Here $n_{\mathrm{T}}$ is the transverse-thrust axis that maximize the quantity and $\lambda_{1,2}$ are the eigenvalues of the transverse linearised sphericity tensor $\mathbf{S}$ defined as
\begin{equation}
    \mathbf{S} = \frac{1}{\sum_i \left| \vec p_{\mathrm{T},i} \right|} \sum_i \frac{1}{\left| \vec p_{\mathrm{T},i} \right|}
    \begin{pmatrix}
        p_{i,x}^2 & p_{i,x} p_{i,y} \\
        p_{i,y} p_{i,x} & p_{i,y}^2
    \end{pmatrix}
\end{equation}
with $i$ summing over the momenta in the final state. In both observables, \Pythia predicts slightly more isotropic events than \Sherpa, which might again be caused by a larger number of MPIs modelled within \Pythia. Comparing the two \Sherpa predictions, we again observe a sizeable $K$ factor and a shift towards more isotropic events. While using a higher value for $\alpha_{S}(M_{\mathrm{Z}})$ in \Pythia as a proxy for the $K$ factor seems to work quite well and gives similar cross sections, the uncertainties from scale variations are significantly smaller at NLO.

\begin{figure}
    \centering
    \includegraphics[width=0.6\linewidth]{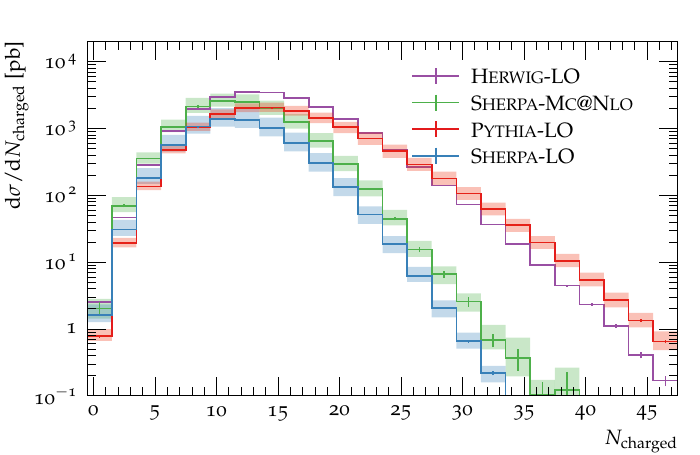}
    \caption{Predictions of charged particle multiplicity $N_\mathrm{charged}$ for events with a jet with $E_{\mathrm{T}} > 6~\text{GeV}$ at the \EIC, comparing to Leading Order simulations by \protect\Herwig, \protect\Pythia and \protect\Sherpa and \protect\MCatNLO-accurate simulations by \protect\Sherpa.}\label{fig:eic-multi}
\end{figure}

As a last observable, in Fig.~\ref{fig:eic-multi} we look at the charged-particle multiplicity in the detector acceptance range $|\eta | < 4$ and see large disagreement between \Pythia and \Sherpa.
Even though hadronisation modelling and MPIs do not play a huge role when studying high-$p_{\mathrm{T}}$ observables, such as jets, they do come into play when studying event structure in more detail.
The result shows less hadrons being generated in \Sherpa than in \Pythia or \Herwig.
We observed that the effect of hadronisation dominates over the MPIs in this observable and it underlines that a careful study of MPIs and hadronisation is necessary to correctly simulate observables like multiplicities.
As discussed before, while the perturbative accuracy is under good control, corrections due to non-perturbative effects rely on data being made available in a format convenient for tuning.

\section{Conclusion}\label{sec:conclusion}

We presented a systematic comparison of the three general-purpose event generators \Herwig, \Pythia, and \Sherpa for jet photoproduction processes.
Starting from a fixed-order baseline, we disentangled the contributions of modelling beyond the hard process — beam remnants, primordial $k_\perp$, ISR and FSR, MPIs, and hadronisation — through a step-by-step comparison.
We used the observable $x_\gamma^\mathrm{obs}$, as it is sensitive to some modelling aspects relevant to photoproduction and therefore of particular interest.
Among the individual effects, the $\gamma \to q\bar{q}$ splitting in the ISR produces a substantial shift of the cross section at $x_\gamma^{\mathrm{obs}} > 0.8$, while MPIs contribute roughly 50\% to the resolved cross section at low $x_\gamma^{\mathrm{obs}}$.
The non-logarithmic photon flux correction accounts for an approximately 10\% reduction in the total cross section in EIC kinematics.

We compared to data for dijet photoproduction analyzed by the \OPAL{} and \ZEUS{} collaborations,
seeing an overall good agreement with the data for the \Sherpa-\MCatNLO{} and the \Pythia{} simulations.
At leading order, to which the \Herwig{} simulations are currently limited, and consistent with \Sherpa{} at leading order, shapes are roughly consistent with the data, however normalizations require a significant $K$-factor.
Furthermore, we presented predictions for the upcoming \EIC{} for inclusive QCD observables and event shapes.
We found significant differences in observables sensitive to the hadronisation and underlying event modelling.
Making more experimental data available in the \Rivet framework would allow for further tuning of MPI-related parameters and should largely resolve such discrepancies.

Finally, we want to emphasise two points. First, because of the variable energy of the incoming photon, photoproduction should provide important constraints on non-perturbative parameters like hadronisation and beam remnants. In this study, we indeed observed large differences between the generators with their different models. Sensitivity to these modellings are barely accessible in fixed-energy collisions at the LHC, hence this information is a complementary handle on these non-perturbative parameters. Second, to fully exploit the constraining power, a precise perturbative baseline is needed. Here the bottleneck is the parton-in-photon PDF sets. The currently available fits were determined more than two decades ago, differ substantially in the gluon content at small $x$ and do not incorporate error estimates as is now standard for proton and nuclear PDF fits. Therefore, a refitting in a global analysis using state-of-the-art methodology is highly desired.

In preparation for the \EIC, photoproduction is the dominant mechanism for hadronic final states, and more work is needed to understand the different regimes and develop a coherent modelling, including the relevant infrastructure in all multi-purpose event generators. Apart from the two points raised above, open questions remain like the transition region between DIS and photoproduction at virtualities of $Q^2 \approx 1 \ \mathrm{GeV}^2$. Progress in this field is a prerequisite of precision phenomenology of hadronic final states at the \EIC.

\section*{Acknowledgements}

We thank the organizers and the staff at the PhysTev 2023 conference, where this study was started.
P.M.\ is supported by the STFC under grant agreement ST/P006744/1.
I.H. has been supported by the Academy of Finland, projects 331545, 361179, and funded as a part of the Center of Excellence in Quark Matter of the Academy of Finland, project 346326. The reported work is associated with the European Research Council project ERC-2018-ADG-835105 YoctoLHC.

\appendix

\section*{Appendix}
\begin{figure}
    \centering
    \begin{tabular}{cc}
        \includegraphics[width=0.4\linewidth]{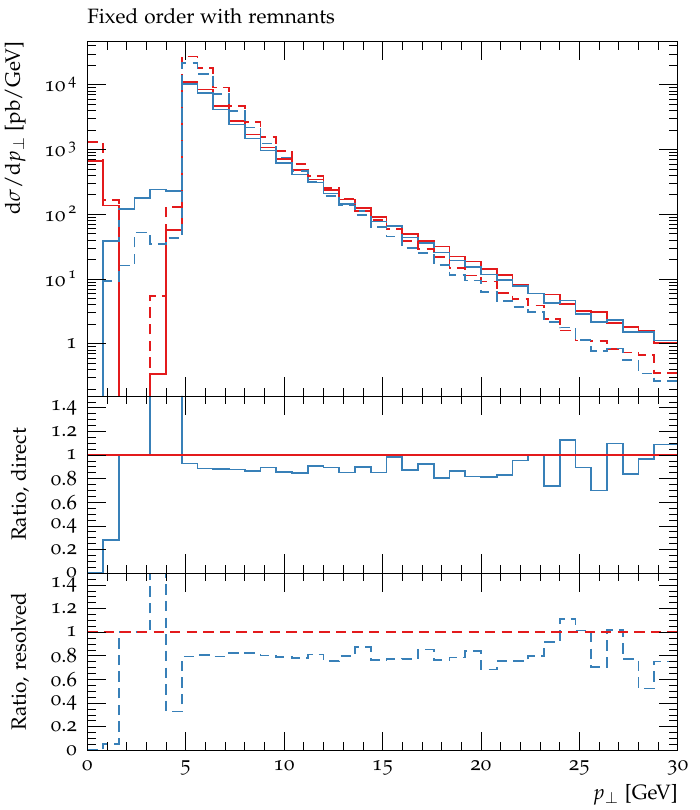} &
        \includegraphics[width=0.4\linewidth]{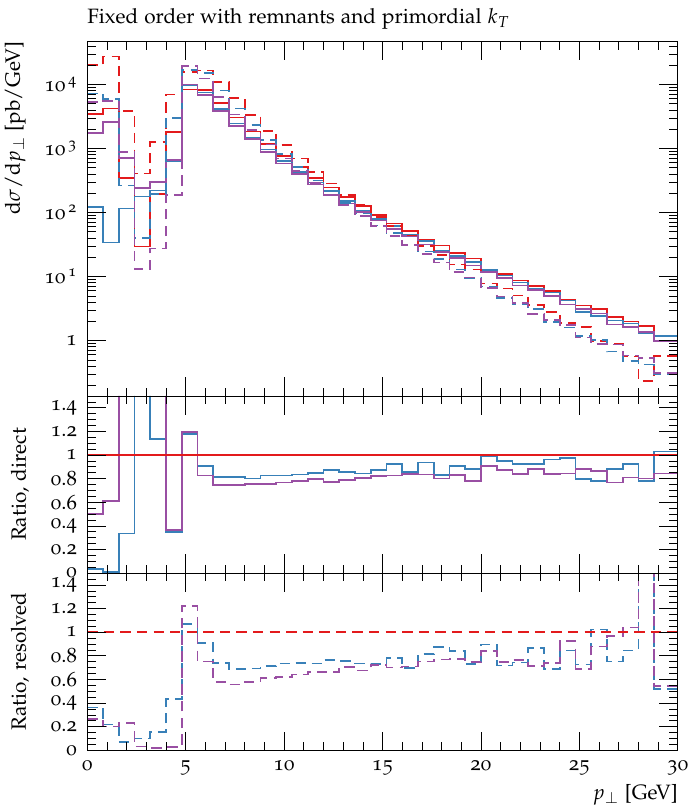} \\
        \includegraphics[width=0.4\linewidth]{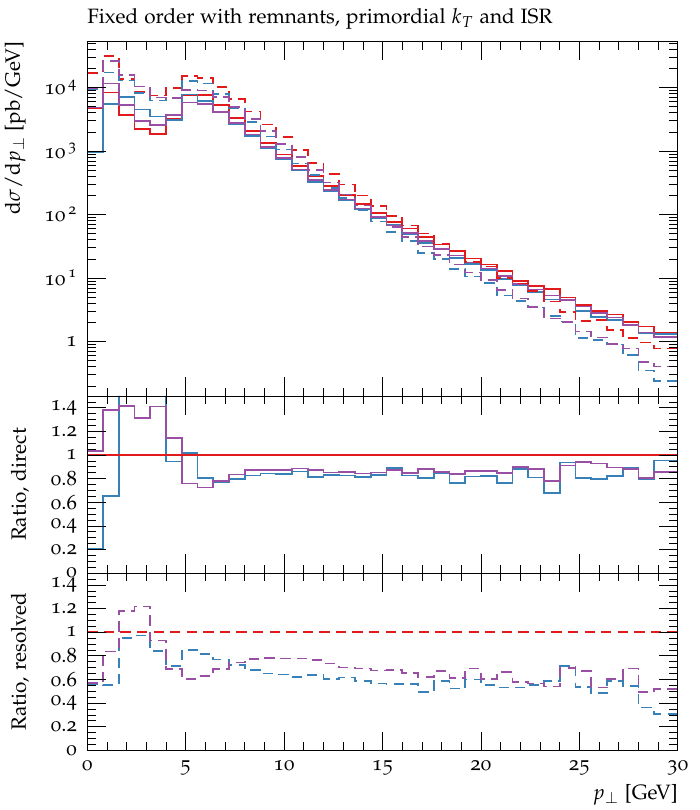} &
        \includegraphics[width=0.4\linewidth]{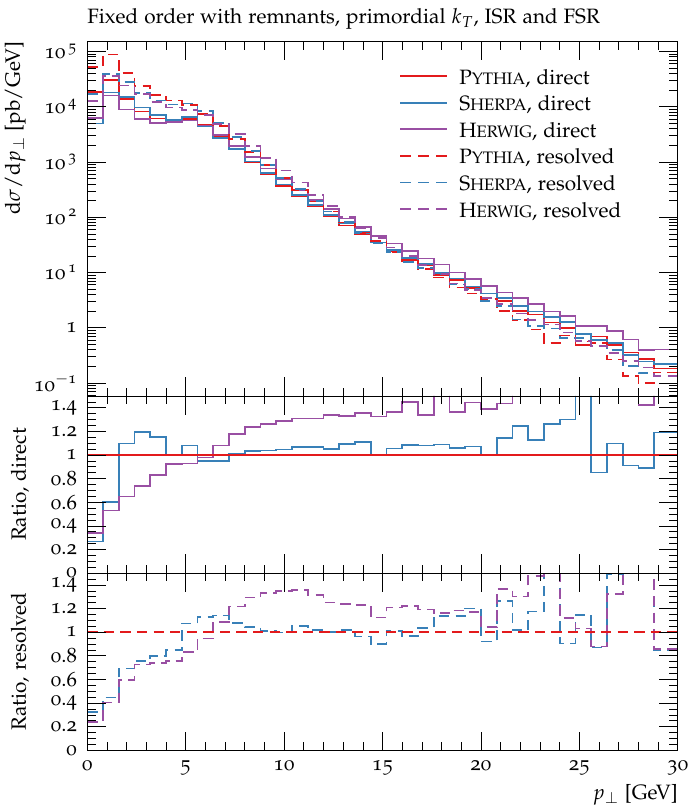} \\
        \includegraphics[width=0.4\linewidth]{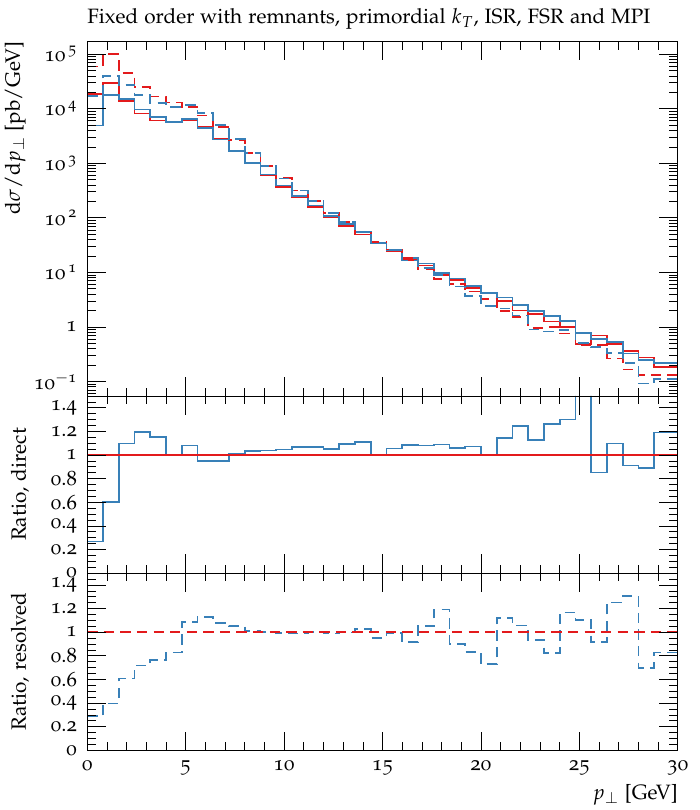} &
        \includegraphics[width=0.4\linewidth]{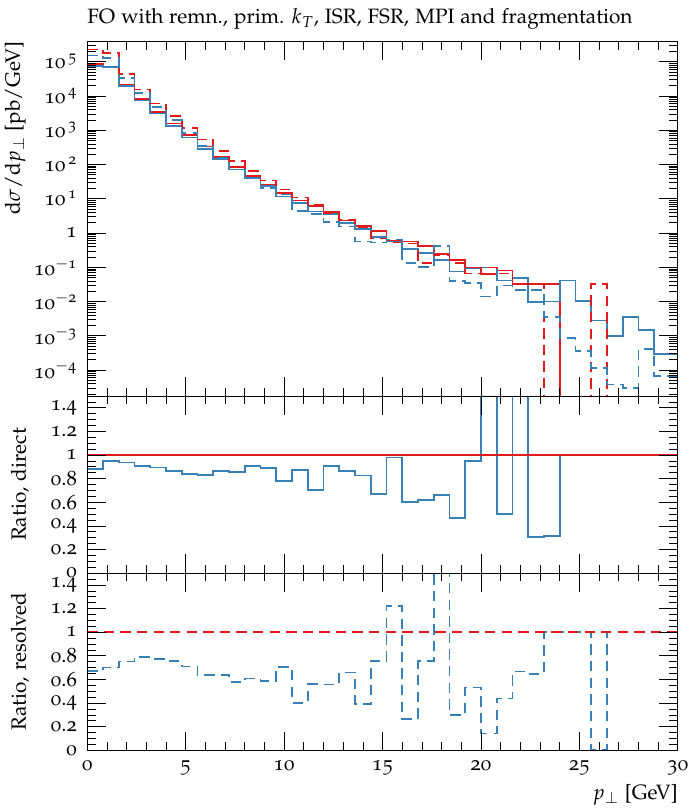}
    \end{tabular}
    \caption{\label{figures:pT-steps}As Fig.~\ref{figures:xg-steps} but for transverse momentum of all final-state particles.}
\end{figure}

\begin{figure}
    \centering
    \begin{tabular}{cc}
        \includegraphics[width=0.4\linewidth]{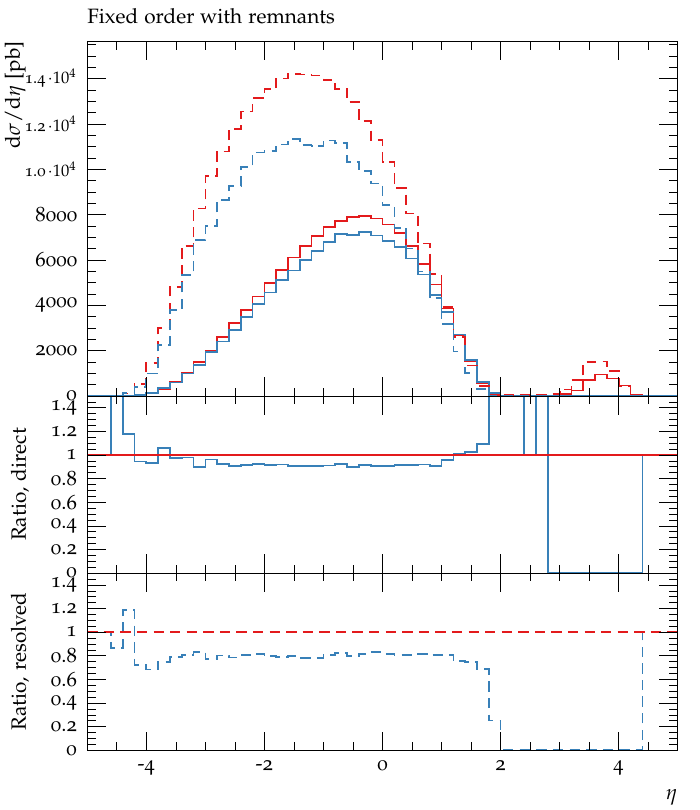} &
        \includegraphics[width=0.4\linewidth]{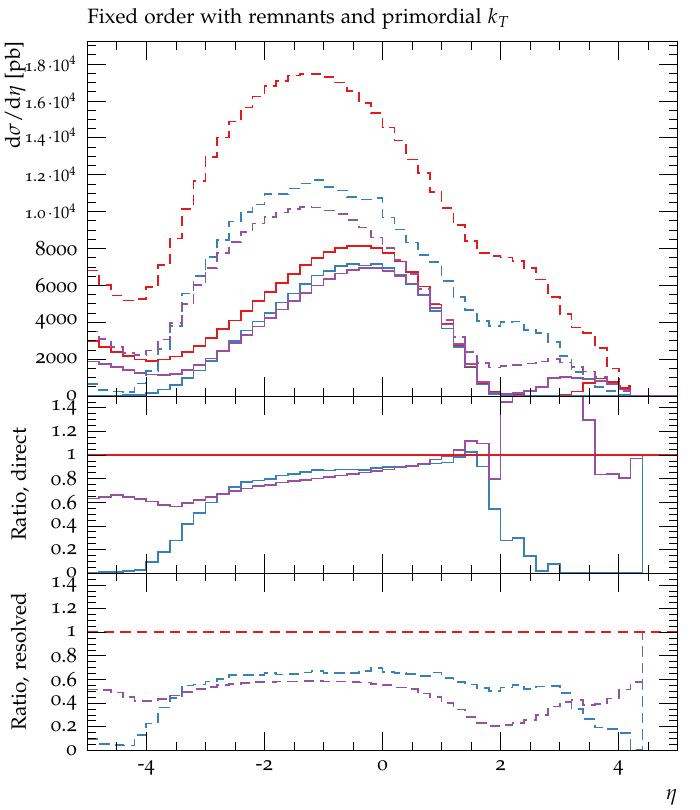} \\
        \includegraphics[width=0.4\linewidth]{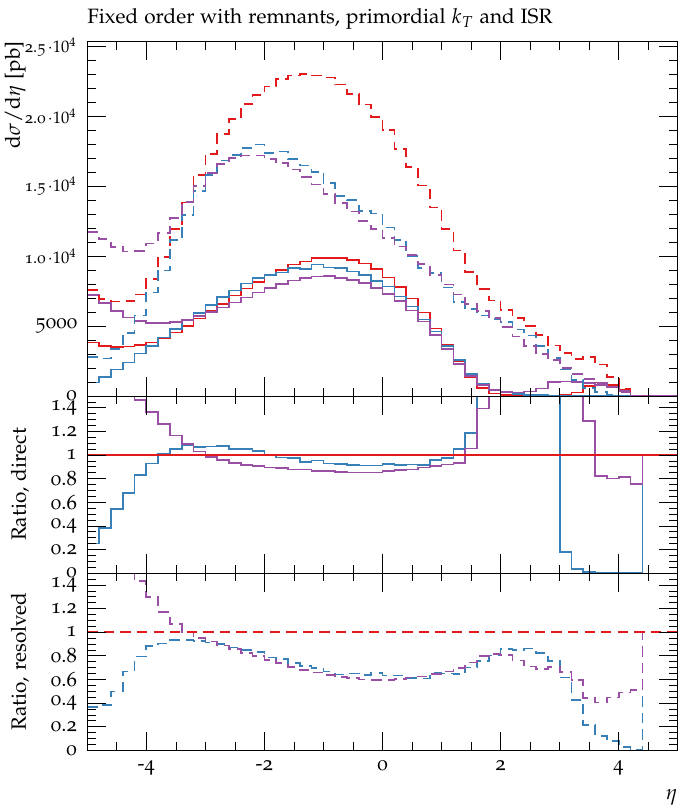} &
        \includegraphics[width=0.4\linewidth]{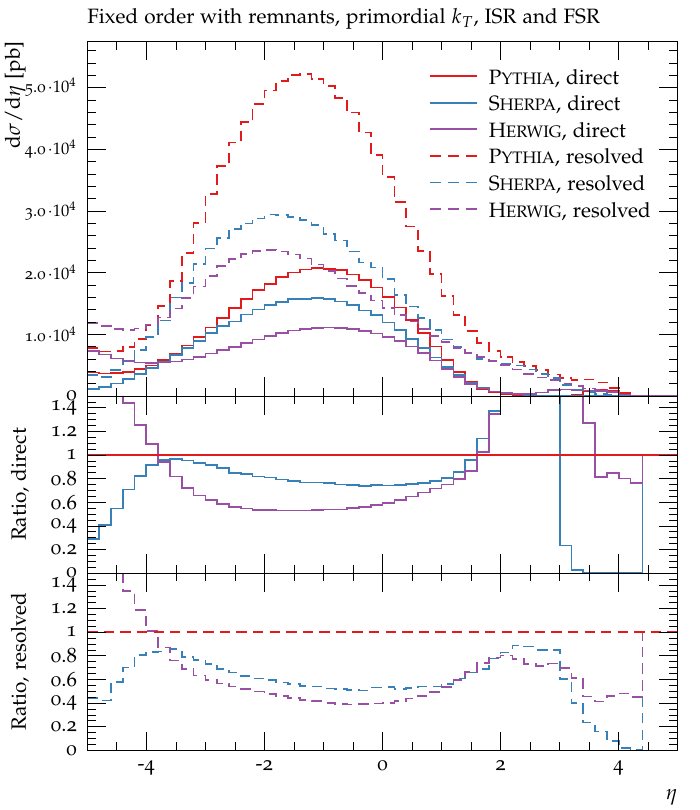} \\
        \includegraphics[width=0.4\linewidth]{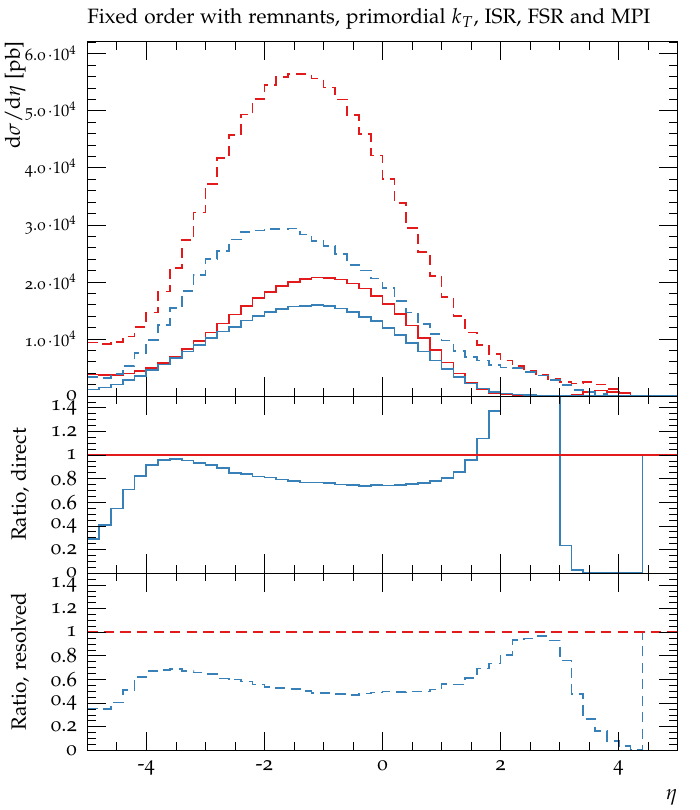} &
        \includegraphics[width=0.4\linewidth]{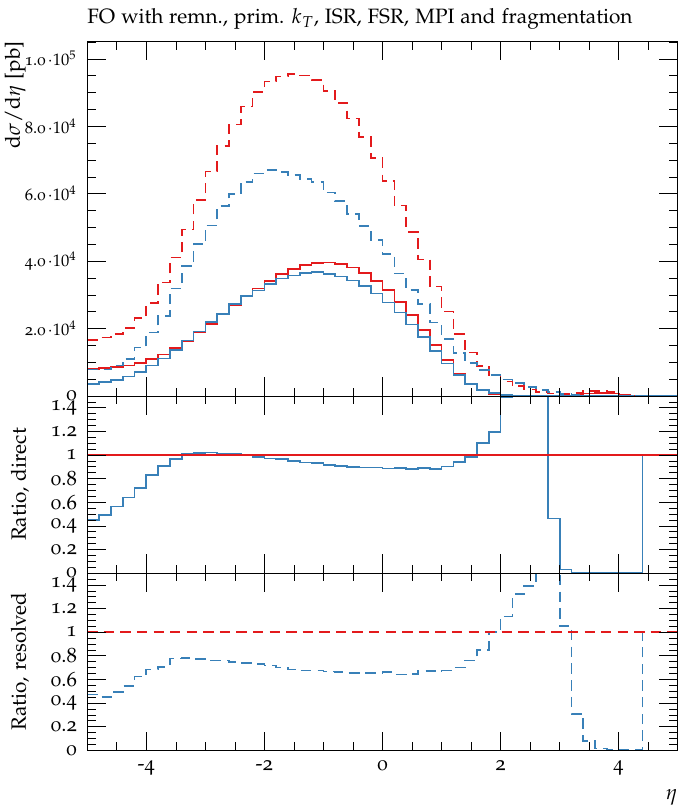}
    \end{tabular}
    \caption{\label{figures:eta-steps}As Fig.~\ref{figures:xg-steps} but for pseudorapidity distribution of all final-state particles.}
\end{figure}

Comparison plots at different stages of event generation for transverse momentum and for pseudorapidity distribution all of particles, shown in Figs.~\ref{figures:pT-steps} and \ref{figures:eta-steps}, respectively.

\bibliographystyle{unsrturl}
\bibliography{refs1,refs2}

@article{Bierlich:2022pfr,
    author = "Bierlich, Christian and others",
    title = "{A comprehensive guide to the physics and usage of PYTHIA 8.3}",
    eprint = "2203.11601",
    archivePrefix = "arXiv",
    primaryClass = "hep-ph",
    reportNumber = "LU-TP 22-16, MCNET-22-04, FERMILAB-PUB-22-227-SCD",
    doi = "10.21468/SciPostPhysCodeb.8",
    month = "3",
    year = "2022",
    journal = "SciPost Phys. Codebases",
    pages = "8",
    publisher = "SciPost"
}

@article{Cornet:2002iy,
    author = "Cornet, F. and Jankowski, P. and Krawczyk, M. and Lorca, A.",
    title = "{A New five flavor LO analysis and parametrization of parton distributions in the real photon}",
    eprint = "hep-ph/0212160",
    archivePrefix = "arXiv",
    reportNumber = "IFT-22-2002, UG-FT-138-02, CAFPE-8-02, DESY-02-118, CERN-TH-2002-362",
    doi = "10.1103/PhysRevD.68.014010",
    journal = "Phys. Rev. D",
    volume = "68",
    pages = "014010",
    year = "2003"
}

@inproceedings{Schuler:1994ft,
    author = "Schuler, Gerhard A. and Sjostrand, Torbjorn",
    title = "{Gamma gamma and gamma P events at high-energies}",
    booktitle = "{Two-Photon Physics from DAPHNE to LEP200 and Beyond}",
    eprint = "hep-ph/9403393",
    archivePrefix = "arXiv",
    reportNumber = "CERN-TH-7193-94",
    month = "3",
    year = "1994"
}

@article{Helenius:2017aqz,
    author = "Helenius, Ilkka",
    editor = "d'Enterria, David and de Roeck, Albert and Mangano, Michelangelo",
    title = "{Photon-photon and photon-hadron processes in Pythia 8}",
    eprint = "1708.09759",
    archivePrefix = "arXiv",
    primaryClass = "hep-ph",
    doi = "10.23727/CERN-Proceedings-2018-001.119",
    journal = "CERN Proc.",
    volume = "1",
    pages = "119",
    year = "2018"
}

@article{ZEUS:2021qzg,
    author = "Abt, I. and others",
    collaboration = "ZEUS",
    title = "{Azimuthal correlations in photoproduction and deep inelastic $ep$ scattering at HERA}",
    eprint = "2106.12377",
    archivePrefix = "arXiv",
    primaryClass = "hep-ex",
    reportNumber = "DESY-21-099",
    doi = "10.1007/JHEP12(2021)102",
    journal = "JHEP",
    volume = "12",
    pages = "102",
    year = "2021"
}

@article{Buccioni:2019sur,
    Author                   = "Buccioni, Federico and Lang, Jean-Nicolas and Lindert, Jonas M. and Maierh{\"o}fer, Philipp and Pozzorini, Stefano and Zhang, Hantian and Zoller, Max F.",
    Title                    = "{OpenLoops 2}",
    eprint = "1907.13071",
    archivePrefix = "arXiv",
    primaryClass = "hep-ph",
    reportNumber = "IPPP/19/62, FR-PHENO-2019-12, PSI-PR-19-15, ZU-TH 37/19",
    doi = "10.1140/epjc/s10052-019-7306-2",
    journal = "Eur. Phys. J. C",
    volume = "79",
    number = "10",
    pages = "866",
    year = "2019"
}

@article{Hoche:2018gti,
      Author                           = "H{\"o}che, Stefan and Kuttimalai, Silvan and Li, Ye",
      title          = "{Hadronic Final States in DIS at NNLO QCD with Parton
                        Showers}",
      journal        = "Phys. Rev.",
      volume         = "D98",
      year           = "2018",
      number         = "11",
      pages          = "114013",
      doi            = "10.1103/PhysRevD.98.114013",
      eprint         = "1809.04192",
      archivePrefix  = "arXiv",
      primaryClass   = "hep-ph",
      reportNumber   = "SLAC-PUB-17319, MCNET-18-24, FERMILAB-PUB-18-510-T,
                        SLAC-PUB-17319 MCNET-18-24",
      SLACcitation   = "%%CITATION = ARXIV:1809.04192;%%"
}

@article{Bellm:2015jjp,
      Author                           = "Bellm, Johannes and others",
      title          = "{Herwig 7.0/Herwig++ 3.0 release note}",
      journal        = "Eur. Phys. J.",
      volume         = "C76",
      year           = "2016",
      number         = "4",
      pages          = "196",
      doi            = "10.1140/epjc/s10052-016-4018-8",
      eprint         = "1512.01178",
      archivePrefix  = "arXiv",
      primaryClass   = "hep-ph",
      reportNumber   = "CERN-PH-TH-2015-289, MAN-HEP-2015-15, IFJPAN-IV-2015-13,
                        KA-TP-18-2015, DCPT-15-142, MCNET-15-28, IPPP-15-71,
                        HERWIG-2015-01",
      SLACcitation   = "%%CITATION = ARXIV:1512.01178;%%"
}

@Article{Andersson:1983ia,
  Title                    = {{Parton Fragmentation and String Dynamics}},
  Author                   = {Andersson, Bo and Gustafson, G. and Ingelman, G. and Sj{\"o}strand, T.},
  Journal                  = {Phys. Rept.},
  Year                     = {1983},
  Pages                    = {31-145},
  Volume                   = {97},

  Doi                      = {10.1016/0370-1573(83)90080-7},
  Slaccitation             = {%%CITATION = PRPLC,97,31;%%}
}

@Article{Bahr:2008pv,
  Title                    = {{Herwig++ Physics and Manual}},
  Author                   = {B{\"a}hr, M. and others},
  Journal                  = {Eur. Phys. J.},
  Year                     = {2008},
  Pages                    = {639-707},
  Volume                   = {C58},

  Archiveprefix            = {arXiv},
  Doi                      = {10.1140/epjc/s10052-008-0798-9},
  Eprint                   = {0803.0883},
  Primaryclass             = {hep-ph},
  Slaccitation             = {%%CITATION = 0803.0883;%%}
}

@Article{Budnev:1974de,
  Title                    = {{The two photon particle production mechanism. Physical problems. Applications. Equivalent photon approximation}},
  Author                   = {Budnev, V. M. and Ginzburg, I. F. and Meledin, G. V. and Serbo, V. G.},
  Journal                  = {Phys. Rept.},
  Year                     = {1974},
  Pages                    = {181-281},
  Volume                   = {15}
}

@Article{Butterworth:2005aq,
  Title                    = {{High energy photoproduction}},
  Author                   = {Butterworth, J. M. and Wing, M.},
  Journal                  = {Rept. Prog. Phys.},
  Year                     = {2005},
  Pages                    = {2773-2828},
  Volume                   = {68},

  Archiveprefix            = {arXiv},
  Doi                      = {10.1088/0034-4885/68/12/R03},
  Eprint                   = {hep-ex/0509018},
  Slaccitation             = {%%CITATION = HEP-EX/0509018;%%}
}

@Article{Collins:1989gx,
  Title                    = {{Factorization of hard processes in QCD}},
  Author                   = {Collins, John C. and Soper, Davison E. and Sterman, George},
  Journal                  = {Adv. Ser. Direct. High Energy Phys.},
  Year                     = {1988},
  Pages                    = {1-91},
  Volume                   = {5},
  Eprint                   = {hep-ph/0409313}
}

@Article{Corcella:2000bw,
  Title                    = {{HERWIG 6: an event generator for hadron emission reactions with interfering gluons (including supersymmetric processes)}},
  Author                   = {Corcella, G. and others},
  Journal                  = {JHEP},
  Year                     = {2001},
  Pages                    = {010},
  Volume                   = {01},

  Archiveprefix            = {arXiv},
  Eprint                   = {hep-ph/0011363},
  Slaccitation             = {%%CITATION = HEP-PH/0011363;%%},
  Url                      = {http://inspirehep.net/search?p=hep-ph/0011363}
}

@article{Bellm:2017idv,
    author = {Bellm, Johannes and Cormier, Kyle and Gieseke, Stefan and Pl\"atzer, Simon and Reuschle, Christian and Richardson, Peter and Webster, Stephen},
    title = "{Top Quark Production and Decay in Herwig 7.1}",
    eprint = "1711.11570",
    archivePrefix = "arXiv",
    primaryClass = "hep-ph",
    reportNumber = "CERN-TH-2017-253, HERWIG-2017-03, IPPP-17-93, KA-TP-41-2017, LU-TP-17-41, MCNET-17-21, UWTHPH-2017-42",
    month = "11",
    year = "2017"
}

@article{Richardson:2018pvo,
    author = "Richardson, Peter and Webster, Stephen",
    title = "{Spin Correlations in Parton Shower Simulations}",
    eprint = "1807.01955",
    archivePrefix = "arXiv",
    primaryClass = "hep-ph",
    reportNumber = "IPPP/18/55, CERN-TH-2018-154, IPPP-18-55, MCNET-18-12",
    doi = "10.1140/epjc/s10052-019-7429-5",
    journal = "Eur. Phys. J. C",
    volume = "80",
    number = "2",
    pages = "83",
    year = "2020"
}

@article{Bewick:2023tfi,
    author = "Bewick, Gavin and others",
    title = "{Herwig 7.3 release note}",
    eprint = "2312.05175",
    archivePrefix = "arXiv",
    primaryClass = "hep-ph",
    reportNumber = "CERN-TH-2023-223, HERWIG-2023-01, KA-TP-28-2023, MCnet-23-19, IPPP/23/66",
    doi = "10.1140/epjc/s10052-024-13211-9",
    journal = "Eur. Phys. J. C",
    volume = "84",
    number = "10",
    pages = "1053",
    year = "2024"
}

@article{Helenius:2024vdj,
    author = "Helenius, Ilkka and Utheim, Marius",
    title = "{Hadron-ion collisions in Pythia and the vector-meson dominance model for photoproduction}",
    eprint = "2406.10403",
    archivePrefix = "arXiv",
    primaryClass = "hep-ph",
    doi = "10.1140/epjc/s10052-024-13543-6",
    journal = "Eur. Phys. J. C",
    volume = "84",
    number = "11",
    pages = "1155",
    year = "2024"
}

@Article{Frixione:2007vw,
  Title                    = {{Matching NLO QCD computations with parton shower simulations: the POWHEG method}},
  Author                   = {Frixione, Stefano and Nason, Paolo and Oleari, Carlo},
  Journal                  = {JHEP},
  Year                     = {2007},
  Pages                    = {070},
  Volume                   = {11},

  Archiveprefix            = {arXiv},
  Eprint                   = {0709.2092},
  Primaryclass             = {hep-ph}
}

@Article{Frixione:2003ei,
  Title                    = {{Matching NLO QCD and parton showers in heavy flavour production}},
  Author                   = {Frixione, S. and Nason, P. and Webber, B. R.},
  Journal                  = {JHEP},
  Year                     = {2003},
  Pages                    = {007},
  Volume                   = {08},

  Eprint                   = {hep-ph/0305252}
}

@Article{Frixione:2002ik,
  Title                    = {{Matching NLO QCD computations and parton shower simulations}},
  Author                   = {Frixione, Stefano and Webber, Bryan R.},
  Journal                  = {JHEP},
  Year                     = {2002},
  Pages                    = {029},
  Volume                   = {06},

  Eprint                   = {hep-ph/0204244}
}

@Article{Gieseke:2003rz,
  Title                    = {{New formalism for QCD parton showers}},
  Author                   = {Gieseke, Stefan and Stephens, P. and Webber, Bryan},
  Journal                  = {JHEP},
  Year                     = {2003},
  Pages                    = {045},
  Volume                   = {12},

  Archiveprefix            = {arXiv},
  Eprint                   = {hep-ph/0310083},
  Slaccitation             = {%%CITATION = HEP-PH/0310083;%%}
}

@article{Gluck:1991jc,
    author = "Gluck, M. and Reya, E. and Vogt, A.",
    title = "{Photonic parton distributions}",
    reportNumber = "DO-TH-91-31",
    doi = "10.1103/PhysRevD.46.1973",
    journal = "Phys. Rev. D",
    volume = "46",
    pages = "1973--1979",
    year = "1992"
}

@Article{Gleisberg:2008fv,
  Title                    = {{Comix, a new matrix element generator}},
  Author                   = {Gleisberg, Tanju and H{\"o}che, Stefan},
  Journal                  = {JHEP},
  Year                     = {2008},
  Pages                    = {039},
  Volume                   = {12},

  Archiveprefix            = {arXiv},
  Doi                      = {10.1088/1126-6708/2008/12/039},
  Eprint                   = {0808.3674},
  Primaryclass             = {hep-ph},
  Slaccitation             = {%%CITATION = 0808.3674;%%},
}

@Article{Gleisberg:2008ta,
  Title                    = {{Event generation with \Sherpa 1.1}},
  Author                   = {Gleisberg, T. and H{\"o}che, S. and Krauss, F. and Sch\"{o}nherr, M. and Schumann, S. and Siegert, F and Winter, J.},
  Journal                  = {JHEP},
  Year                     = {2009},
  Pages                    = {007},
  Volume                   = {02},

  Archiveprefix            = {arXiv},
  Doi                      = {10.1088/1126-6708/2009/02/007},
  Eprint                   = {0811.4622},
  Primaryclass             = {hep-ph},
  Slaccitation             = {%%CITATION = 0811.4622;%%}
}

@Article{Gleisberg:2007md,
  Title                    = {{Automating dipole subtraction for QCD NLO calculations}},
  Author                   = {Gleisberg, Tanju and Krauss, Frank},
  Journal                  = {Eur. Phys. J.},
  Year                     = {2008},
  Pages                    = {501-523},
  Volume                   = {C53},

  Archiveprefix            = {arXiv},
  Eprint                   = {0709.2881},
  Primaryclass             = {hep-ph}
}

@Article{Hoeche:2011fd,
  Title                    = {{A critical appraisal of NLO+PS matching methods}},
  Author                   = {H{\"o}che, Stefan and Krauss, Frank and Sch{\"o}nherr, Marek and Siegert, Frank},
  Journal                  = {JHEP},
  Year                     = {2012},
  Pages                    = {049},
  Volume                   = {09},

  Archiveprefix            = {arXiv},
  Eprint                   = {1111.1220},
  Primaryclass             = {hep-ph}
}

@Article{Krauss:2001iv,
  Title                    = {{AMEGIC++ 1.0: A Matrix Element Generator In C++}},
  Author                   = {Frank Krauss and Ralf Kuhn and Gerhard Soff},
  Journal                  = {JHEP},
  Year                     = {2002},
  Pages                    = {044},
  Volume                   = {02},

  Eprint                   = {hep-ph/0109036}
}

@Article{Lavesson:2008ah,
  Title                    = {{Extending CKKW-merging to one-loop matrix elements}},
  Author                   = {Lavesson, Nils and L{\"o}nnblad, Leif},
  Journal                  = {JHEP},
  Year                     = {2008},
  Pages                    = {070},
  Volume                   = {12},

  Archiveprefix            = {arXiv},
  Doi                      = {10.1088/1126-6708/2008/12/070},
  Eprint                   = {0811.2912},
  Primaryclass             = {hep-ph},
  Slaccitation             = {%%CITATION = 0811.2912;%%}
}

@Article{Platzer:2012bs,
  Title                    = {{Controlling inclusive cross sections in parton shower + matrix element merging}},
  Author                   = {Pl{\"a}tzer, Simon},
  Journal                  = {JHEP},
  Year                     = {2013},
  Pages                    = {114},
  Volume                   = {08},

  Archiveprefix            = {arXiv},
  Doi                      = {10.1007/JHEP08(2013)114},
  Eprint                   = {1211.5467},
  Primaryclass             = {hep-ph},
  Reportnumber             = {DESY-12-215, MCNET-12-13},
  Slaccitation             = {%%CITATION = ARXIV:1211.5467;%%}
}

@Article{Platzer:2011bc,
  Title                    = {{Dipole Showers and Automated NLO Matching in Herwig++}},
  Author                   = {Pl{\"a}tzer, Simon and Gieseke, Stefan},
  Journal                  = {Eur.Phys.J.},
  Year                     = {2012},
  Pages                    = {2187},
  Volume                   = {C72},

  Archiveprefix            = {arXiv},
  Doi                      = {10.1140/epjc/s10052-012-2187-7},
  Eprint                   = {1109.6256},
  Primaryclass             = {hep-ph},
  Reportnumber             = {DESY-11-162, KA-TP-24-2011, HERWIG-11-01, MCNET-11-24},
  Slaccitation             = {%%CITATION = ARXIV:1109.6256;%%}
}

@Article{Platzer:2009jq,
  Title                    = {{Coherent Parton Showers with Local Recoils}},
  Author                   = {Pl{\"a}tzer, Simon and Gieseke, Stefan},
  Journal                  = {JHEP},
  Year                     = {2011},
  Pages                    = {024},
  Volume                   = {01},

  Archiveprefix            = {arXiv},
  Doi                      = {10.1007/JHEP01(2011)024},
  Eprint                   = {0909.5593},
  Primaryclass             = {hep-ph},
  Slaccitation             = {%%CITATION = 0909.5593;%%}
}

@Article{Schuler:1996en,
  Title                    = {{A scenario for high-energy $\gamma \gamma$ interactions}},
  Author                   = {Schuler, Gerhard A. and Sj{\"o}strand, Torbj{\"o}rn},
  Journal                  = {Z. Phys.},
  Year                     = {1997},
  Pages                    = {677-688},
  Volume                   = {C73},

  Archiveprefix            = {arXiv},
  Doi                      = {10.1007/s002880050359},
  Eprint                   = {hep-ph/9605240},
  Slaccitation             = {%%CITATION = HEP-PH/9605240;%%},
  Url                      = {http://inspirehep.net/search?p=hep-ph/9605240}
}

@Article{Schumann:2007mg,
  Title                    = {{A parton shower algorithm based on Catani-Seymour dipole factorisation}},
  Author                   = {Schumann, Steffen and Krauss, Frank},
  Journal                  = {JHEP},
  Year                     = {2008},
  Pages                    = {038},
  Volume                   = {03},

  Archiveprefix            = {arXiv},
  Eprint                   = {0709.1027},
  Primaryclass             = {hep-ph}
}

@Article{Sjostrand:1987su,
  Title                    = {{A multiple-interaction model for the event structure in hadron collisions}},
  Author                   = {Torbj{\"o}rn Sj{\"o}strand and Maria van Zijl},
  Journal                  = {Phys. Rev.},
  Year                     = {1987},
  Pages                    = {2019},
  Volume                   = {D36}
}

@Article{Webber:1983if,
  Title                    = {{A QCD model for jet fragmentation including soft gluon interference}},
  Author                   = {Bryan R. Webber},
  Journal                  = {Nucl. Phys.},
  Year                     = {1984},
  Pages                    = {492},
  Volume                   = {B238}
}

@inproceedings{Hoche:2014rga,
      Author                           = "H{\"o}che, Stefan",
      title          = "{Introduction to parton-shower event generators}",
      booktitle      = "{Proceedings, Theoretical Advanced Study Institute in
                        Elementary Particle Physics: Journeys Through the
                        Precision Frontier: Amplitudes for Colliders (TASI 2014):
                        Boulder, Colorado, June 2-27, 2014}",
      year           = "2015",
      pages          = "235-295",
      doi            = "10.1142/9789814678766_0005",
      eprint         = "1411.4085",
      archivePrefix  = "arXiv",
      primaryClass   = "hep-ph",
      reportNumber   = "SLAC-PUB-16160",
      SLACcitation   = "%%CITATION = ARXIV:1411.4085;%%"
}

@article{Bothmann:2019yzt,
      Author                           = "Bothmann, Enrico and others",
      title          = "{Event Generation with Sherpa 2.2}",
      journal        = "SciPost Phys.",
      volume         = "7",
      year           = "2019",
      number         = "3",
      pages          = "034",
      doi            = "10.21468/SciPostPhys.7.3.034",
      eprint         = "1905.09127",
      archivePrefix  = "arXiv",
      primaryClass   = "hep-ph",
      reportNumber   = "FERMILAB-PUB-19-218-T, SLAC-PUB-17433, IPPP/19/42,
                        MCNET-19-11",
      SLACcitation   = "%%CITATION = ARXIV:1905.09127;%%"
}

@article{Bellm:2017ktr,
      Author                           = "Bellm, Johannes and Gieseke, Stefan and Plätzer, Simon",
      title          = "{Merging NLO Multi-jet Calculations with Improved
                        Unitarization}",
      journal        = "Eur. Phys. J.",
      volume         = "C78",
      year           = "2018",
      number         = "3",
      pages          = "244",
      doi            = "10.1140/epjc/s10052-018-5723-2",
      eprint         = "1705.06700",
      archivePrefix  = "arXiv",
      primaryClass   = "hep-ph",
      reportNumber   = "IPPP-17-39, KA-TP-20-2017, MAN-HEP-2017-07,
                        UWTHPH-2017-9, MCNET-17-07, HERWIG-2017-01",
      SLACcitation   = "%%CITATION = ARXIV:1705.06700;%%"
}

@article{ZEUS:2001zoq,
    author = "Chekanov, S. and others",
    collaboration = "ZEUS",
    title = "{Dijet photoproduction at HERA and the structure of the photon}",
    eprint = "hep-ex/0112029",
    archivePrefix = "arXiv",
    reportNumber = "DESY-01-220",
    doi = "10.1007/s100520200936",
    journal = "Eur. Phys. J. C",
    volume = "23",
    pages = "615--631",
    year = "2002"
}

@article{OPAL:2003hoh,
    author = "Abbiendi, G. and others",
    collaboration = "OPAL",
    title = "{Dijet production in photon-photon collisions at s(ee)**(1/2) from 189-GeV to 209-GeV}",
    eprint = "hep-ex/0301013",
    archivePrefix = "arXiv",
    reportNumber = "CERN-EP-2002-093",
    doi = "10.1140/epjc/s2003-01360-8",
    journal = "Eur. Phys. J. C",
    volume = "31",
    pages = "307--325",
    year = "2003"
}

@misc{sherpa-manual,
  title = {{Sherpa Project webpage}},
  howpublished = {\url{https://sherpa-team.gitlab.io/}},
  note = {Accessed: 13 Oct 2023}
}

@article{Schuler:1993wr,
    author = "Schuler, Gerhard A. and Sjostrand, Torbjorn",
    title = "{Hadronic diffractive cross-sections and the rise of the total cross-section}",
    reportNumber = "CERN-TH-6837-93",
    doi = "10.1103/PhysRevD.49.2257",
    journal = "Phys. Rev. D",
    volume = "49",
    pages = "2257--2267",
    year = "1994"
}

@article{Chahal:2022rid,
    author = "Chahal, Gurpreet Singh and Krauss, Frank",
    title = "{Cluster Hadronisation in Sherpa}",
    eprint = "2203.11385",
    archivePrefix = "arXiv",
    primaryClass = "hep-ph",
    reportNumber = "IPPP/22/14",
    doi = "10.21468/SciPostPhys.13.2.019",
    journal = "SciPost Phys.",
    volume = "13",
    number = "2",
    pages = "019",
    year = "2022"
}

@Article{Schuler1995,
  author    = {Gerhard A. Schuler and Torbjörn Sjöstrand},
  title     = {Low- and high-mass components of the photon distribution functions},
  doi       = {10.1007/bf01565260},
  eprint    = {hep-ph/9503384},
  number    = {4},
  pages     = {607--623},
  volume    = {68},
  journal   = {Zeitschrift für Physik C Particles and Fields},
  month     = dec,
  publisher = {Springer Science and Business Media {LLC}},
  year      = {1995},
}

@Article{Schuler1996a,
  author        = {Gerhard A. Schuler and Torbjörn Sjöstrand},
  title         = {{Parton Distributions of the Virtual Photon}},
  doi           = {10.1016/0370-2693(96)00265-1},
  eprint        = {hep-ph/9601282},
  archiveprefix = {arXiv},
  journal       = {Phys.Lett.B376:193-200,1996},
  month         = jan,
  primaryclass  = {hep-ph},
  year          = {1996},
}

@article{Frixione:1993yw,
    author = "Frixione, Stefano and Mangano, Michelangelo L. and Nason, Paolo and Ridolfi, Giovanni",
    title = "{Improving the Weizsacker-Williams approximation in electron - proton collisions}",
    eprint = "hep-ph/9310350",
    archivePrefix = "arXiv",
    reportNumber = "CERN-TH-7032-93, GEF-TH-18-93",
    doi = "10.1016/0370-2693(93)90823-Z",
    journal = "Phys. Lett. B",
    volume = "319",
    pages = "339--345",
    year = "1993"
}

@article{PDF4LHCWorkingGroup:2022cjn,
    author = "Ball, Richard D. and others",
    collaboration = "PDF4LHC Working Group",
    title = "{The PDF4LHC21 combination of global PDF fits for the LHC Run III}",
    eprint = "2203.05506",
    archivePrefix = "arXiv",
    primaryClass = "hep-ph",
    reportNumber = "Edinburgh 2021/31, FERMILAB-PUB-22-121-QIS-SCD-T, MSUHEP-22-010,
  Nikhef 2021-033, SMU-HEP-22-01",
    doi = "10.1088/1361-6471/ac7216",
    journal = "J. Phys. G",
    volume = "49",
    number = "8",
    pages = "080501",
    year = "2022"
}

@article{AbdulKhalek:2021gbh,
    author = "Abdul Khalek, R. and others",
    title = "{Science Requirements and Detector Concepts for the Electron-Ion Collider}: {EIC Yellow Report}",
    eprint = "2103.05419",
    archivePrefix = "arXiv",
    primaryClass = "physics.ins-det",
    reportNumber = "BNL-220990-2021-FORE, JLAB-PHY-21-3198, LA-UR-21-20953",
    doi = "10.1016/j.nuclphysa.2022.122447",
    journal = "Nucl. Phys. A",
    volume = "1026",
    pages = "122447",
    year = "2022"
}

@article{Hoeche:2023gme,
    author = "Hoeche, Stefan and Krauss, Frank and Meinzinger, Peter",
    title = "{Resolved Photons in Sherpa}",
    eprint = "2310.18674",
    archivePrefix = "arXiv",
    primaryClass = "hep-ph",
    journal = "Eur. Phys. J. C",
    month = "10",
    date = "2024/02/21",
    doi = "10.1140/epjc/s10052-024-12551-w",
    number = "2",
    pages = "178",
    volume = "84",
    year = "2024"
}

@article{Meinzinger:2023xuf,
    author = "Meinzinger, Peter and Krauss, Frank",
    title = "{Hadron-level NLO predictions for QCD observables in photo-production at the Electron-Ion Collider}",
    eprint = "2311.14571",
    archivePrefix = "arXiv",
    primaryClass = "hep-ph",
    doi = "10.1103/PhysRevD.109.034037",
    journal = "Phys. Rev. D",
    volume = "109",
    number = "3",
    pages = "034037",
    year = "2024"
}

@article{ParticleDataGroup:2020ssz,
    author = "Zyla, P. A. and others",
    collaboration = "Particle Data Group",
    title = "{Review of Particle Physics}",
    doi = "10.1093/ptep/ptaa104",
    journal = "PTEP",
    volume = "2020",
    number = "8",
    pages = "083C01",
    year = "2020"
}

@inproceedings{Seymour:2013ega,
    author = "Seymour, Michael H. and Marx, Marilyn",
    title = "{Monte Carlo Event Generators}",
    booktitle = "{69th Scottish Universities Summer School in Physics}: {LHC Physics}",
    eprint = "1304.6677",
    archivePrefix = "arXiv",
    primaryClass = "hep-ph",
    reportNumber = "MCNET-13-05",
    doi = "10.1007/978-3-319-05362-2_8",
    pages = "287--319",
    month = "4",
    year = "2013"
}

@book{Bartalini:2018qje,
    editor = "Bartalini, Paolo and Gaunt, Jonathan Richard",
    title = "{Multiple Parton Interactions at the LHC}",
    doi = "10.1142/10646",
    isbn = "978-981-322-775-0, 978-981-322-777-4",
    publisher = "WSP",
    volume = "29",
    year = "2019"
}

@article{Sjostrand:1993hi,
    author = "Sjostrand, Torbjorn and Khoze, Valery A.",
    title = "{On Color rearrangement in hadronic W+ W- events}",
    eprint = "hep-ph/9310242",
    archivePrefix = "arXiv",
    reportNumber = "CERN-TH-7011-93, DTP-93-74",
    doi = "10.1007/BF01560244",
    journal = "Z. Phys. C",
    volume = "62",
    pages = "281--310",
    year = "1994"
}

@article{Christiansen:2015yqa,
    author = "Christiansen, Jesper R. and Skands, Peter Z.",
    title = "{String Formation Beyond Leading Colour}",
    eprint = "1505.01681",
    archivePrefix = "arXiv",
    primaryClass = "hep-ph",
    reportNumber = "COEPP-MN-15-1, LU-TP-15-16, MCNET-15-09, COEPP-MN-15-1, LU-TP-15-16, MCNET-15-09",
    doi = "10.1007/JHEP08(2015)003",
    journal = "JHEP",
    volume = "08",
    pages = "003",
    year = "2015"
}

@article{Bierlich:2018xfw,
    author = {Bierlich, Christian and Gustafson, G\"osta and L\"onnblad, Leif and Shah, Harsh},
    title = "{The Angantyr model for Heavy-Ion Collisions in PYTHIA8}",
    eprint = "1806.10820",
    archivePrefix = "arXiv",
    primaryClass = "hep-ph",
    reportNumber = "LU-TP-18-19, LU-TP 18-19, MCnet-18-12",
    doi = "10.1007/JHEP10(2018)134",
    journal = "JHEP",
    volume = "10",
    pages = "134",
    year = "2018"
}

@article{Cabouat:2017rzi,
    author = {Cabouat, Baptiste and Sj\"ostrand, Torbj\"orn},
    title = "{Some Dipole Shower Studies}",
    eprint = "1710.00391",
    archivePrefix = "arXiv",
    primaryClass = "hep-ph",
    reportNumber = "MCNET-17-14, LU-TP-17-28",
    doi = "10.1140/epjc/s10052-018-5645-z",
    journal = "Eur. Phys. J. C",
    volume = "78",
    number = "3",
    pages = "226",
    year = "2018"
}

@article{Helenius:2019gbd,
    author = "Helenius, Ilkka and Rasmussen, Christine O.",
    title = "{Hard diffraction in photoproduction with Pythia 8}",
    eprint = "1901.05261",
    archivePrefix = "arXiv",
    primaryClass = "hep-ph",
    reportNumber = "LU TP 19-06, MCNET-19-01",
    doi = "10.1140/epjc/s10052-019-6914-1",
    journal = "Eur. Phys. J. C",
    volume = "79",
    number = "5",
    pages = "413",
    year = "2019"
}

@article{Knobbe:2023ehi,
    author = "Knobbe, Max and Reichelt, Daniel and Schumann, Steffen",
    title = "{(N)NLO+NLL\textquoteright{} accurate predictions for plain and groomed 1-jettiness in neutral current DIS}",
    eprint = "2306.17736",
    archivePrefix = "arXiv",
    primaryClass = "hep-ph",
    reportNumber = "MCNET-23-07, IPPP/23/32",
    doi = "10.1007/JHEP09(2023)194",
    journal = "JHEP",
    volume = "09",
    pages = "194",
    year = "2023"
}

@article{Bertone:2022hig,
    author = "Bertone, Valerio and Prestel, Stefan",
    title = "{Combining N3LO QCD calculations and parton showers for hadronic collision events}",
    eprint = "2202.01082",
    archivePrefix = "arXiv",
    primaryClass = "hep-ph",
    reportNumber = "LU-TP-22-06",
    month = "2",
    year = "2022"
}

@article{Budnev:1975poe,
    author = "Budnev, V. M. and Ginzburg, I. F. and Meledin, G. V. and Serbo, V. G.",
    title = "{The Two photon particle production mechanism. Physical problems. Applications. Equivalent photon approximation}",
    doi = "10.1016/0370-1573(75)90009-5",
    journal = "Phys. Rept.",
    volume = "15",
    pages = "181--281",
    year = "1975"
}

@article{OPAL:2006pyk,
    author = "Abbiendi, G. and others",
    collaboration = "OPAL",
    title = "{Inclusive production of charged hadrons in photon-photon collisions}",
    eprint = "hep-ex/0612045",
    archivePrefix = "arXiv",
    reportNumber = "CERN-PH-EP-2006-038",
    doi = "10.1016/j.physletb.2007.06.001",
    journal = "Phys. Lett. B",
    volume = "651",
    pages = "92--101",
    year = "2007"
}

@article{Bauer:1977iq,
    author = "Bauer, T. H. and Spital, R. D. and Yennie, D. R. and Pipkin, F. M.",
    title = "{The Hadronic Properties of the Photon in High-Energy Interactions}",
    reportNumber = "PRINT-77-0549 (HARVARD)",
    doi = "10.1103/RevModPhys.50.261",
    journal = "Rev. Mod. Phys.",
    volume = "50",
    pages = "261",
    year = "1978",
    note = "[Erratum: Rev.Mod.Phys. 51, 407 (1979)]"
}

@article{Klasen:2002xb,
    author = "Klasen, Michael",
    title = "{Theory of hard photoproduction}",
    eprint = "hep-ph/0206169",
    archivePrefix = "arXiv",
    reportNumber = "DESY-02-086",
    doi = "10.1103/RevModPhys.74.1221",
    journal = "Rev. Mod. Phys.",
    volume = "74",
    pages = "1221--1282",
    year = "2002"
}

@article{Krawczyk:2000mf,
    author = "Krawczyk, Maria and Zembrzuski, Andrzej and Staszel, Magdalena",
    title = "{Survey of present data on photon structure functions and resolved photon processes}",
    eprint = "hep-ph/0011083",
    archivePrefix = "arXiv",
    reportNumber = "IFT-99-15",
    doi = "10.1016/S0370-1573(00)00105-8",
    journal = "Phys. Rept.",
    volume = "345",
    pages = "265--450",
    year = "2001"
}

@article{Nisius:1999cv,
    author = "Nisius, Richard",
    title = "{The Photon structure from deep inelastic electron photon scattering}",
    eprint = "hep-ex/9912049",
    archivePrefix = "arXiv",
    doi = "10.1016/S0370-1573(99)00115-5",
    journal = "Phys. Rept.",
    volume = "332",
    pages = "165--317",
    year = "2000"
}

@article{Walsh:1973mz,
    author = "Walsh, T. F. and Zerwas, Peter M.",
    title = "{Two photon processes in the parton model}",
    reportNumber = "DESY-72-77",
    doi = "10.1016/0370-2693(73)90520-0",
    journal = "Phys. Lett. B",
    volume = "44",
    pages = "195--198",
    year = "1973"
}

@article{Abramowicz:1993xb,
    author = "Abramowicz, H. and Krawczyk, M. and Charchula, K. and Levy, A. and Maor, U.",
    title = "{Parton distributions in the photon}",
    doi = "10.1142/S0217751X93000394",
    journal = "Int. J. Mod. Phys. A",
    volume = "8",
    pages = "1005--1040",
    year = "1993"
}

@article{DeWitt:1978wn,
    author = "DeWitt, R. J. and Jones, L. M. and Sullivan, J. D. and Willen, D. E. and Wyld, Jr., H. W.",
    title = "{Anomalous Components of the Photon Structure Functions}",
    reportNumber = "ILL-TH-78-54",
    doi = "10.1103/PhysRevD.19.2046",
    journal = "Phys. Rev. D",
    volume = "19",
    pages = "2046",
    year = "1979",
    note = "[Erratum: Phys.Rev.D 20, 1751 (1979)]"
}

@article{Frazer:1979gc,
    author = "Frazer, William R. and Gunion, John F.",
    title = "{Tests of QCD in Two Photon Collisions and High p(t) Photon Production}",
    reportNumber = "UCD-78-5, UCSD-10P10-199",
    doi = "10.1103/PhysRevD.20.147",
    journal = "Phys. Rev. D",
    volume = "20",
    pages = "147",
    year = "1979"
}

@article{Gluck:1983bh,
    author = "Gluck, M. and Grassie, K. and Reya, E.",
    title = "{Detailed {QCD} Analysis of the Photon Structure Function}",
    reportNumber = "DO-TH 83/23",
    doi = "10.1103/PhysRevD.30.1447",
    journal = "Phys. Rev. D",
    volume = "30",
    pages = "1447",
    year = "1984"
}

@article{Schuler:1992dt,
    author = "Schuler, Gerhard A. and Sjostrand, Torbjorn",
    title = "{The hadronic properties of the photon in gamma p interactions}",
    reportNumber = "CERN-TH-6718-92",
    doi = "10.1016/0370-2693(93)90766-B",
    journal = "Phys. Lett. B",
    volume = "300",
    pages = "169--174",
    year = "1993"
}

\end{document}